\documentclass[aps,superscriptaddress,reprint]{revtex4-2}
\usepackage{graphicx}
\usepackage{dcolumn}
\usepackage{bm}
\usepackage{amssymb}
\usepackage{amsmath}
\usepackage{hyperref}
\usepackage{epsf}
\usepackage{color}
\draft

\begin{document}
\preprint{CeRhIn5/QCP}

\title{Critical-point anomalies in doped CeRhIn$_5$}
\author{R. Mathew Roy}
\email{renjith.mathew-roy@pi1.uni-stuttgart.de}
\affiliation{1.~Physikalisches Institut, Universit\"at Stuttgart, Pfaffenwaldring 57,
70569 Stuttgart, Germany}
\author{S. Pal}
\affiliation{1.~Physikalisches Institut, Universit\"at Stuttgart, Pfaffenwaldring 57,
70569 Stuttgart, Germany}
 \author{R. Yang}
\affiliation{1.~Physikalisches Institut, Universit\"at Stuttgart, Pfaffenwaldring 57,
70569 Stuttgart, Germany}
  \author{S. Roh}
\affiliation{1.~Physikalisches Institut, Universit\"at Stuttgart, Pfaffenwaldring 57,
70569 Stuttgart, Germany}
 \author{S. Shin}
 \affiliation{Laboratory for Multiscale Materials Experiments, Paul Scherrer Institut, CH-5232 Villigen PSI, Switzerland}
 \author{T. B. Park}
 \affiliation{Centre for Quantum Materials and Superconductivity (CQMS), Sungkyunkwan University, 16419 Suwon, Republic of Korea}
 \author{T. Park}
 \affiliation{Centre for Quantum Materials and Superconductivity (CQMS), Sungkyunkwan University, 16419 Suwon, Republic of Korea}
 \author{M. Dressel}
\affiliation{1.~Physikalisches Institut, Universit\"at Stuttgart, Pfaffenwaldring 57,
70569 Stuttgart, Germany}

\begin{abstract}

The heavy-fermion compound CeRhIn$_5$  can be tuned through a quantum critical point,
when In is partially replaced by Sn. This way additional charge carriers are introduced and the antiferromagnetic order is gradually suppressed to zero temperature.
Here we investigate the temperature-dependent optical properties of CeRh(In$_{1-x}$Sn$_x$)$_5$ single crystals for
$x = 4.4\%$, 6.9\% and 9.8\%. With increasing Sn concentration the infrared conductivity reveals a clear enhancement of the $c$-$f$ hybridization strength. At low temperatures we observed a non-Fermi-liquid behavior in the frequency dependence of the scattering rate and effective mass in all three compounds. In addition, below a characteristic temperature $T^* \approx 10$~K, the temperature dependent resistivity $\rho(T)$ follows a $\log T$ behavior, typical for a non-Fermi liquid. The temperature-dependent magnetization also exhibits anomalous behavior below $T^*$. Our investigation reveal that below $T^*$ the system shows a pronounced non-Fermi-liquid behavior and $T^*$ monotonically increases as
the quantum critical point is approached.

\end{abstract}

\date{\today}%
\maketitle

\section{introduction}
Quantum phase transitions in strongly correlated electron systems have been studied intensely for decades,
where most candidates are found among high-$T_c$ superconductors, organic conductors, heavy fermions, etc. \cite{Si2010,Paschen2021,Dressel2011,*Dressel2020,Lohneysen2007}. However, the unconventional response of systems near the quantum critical point is rather complicated and yet not fully understood \cite{Kuga2018}.  In this context, the heavy-fermion metals are an excellent playground for exploring quantum criticality and non-Fermi liquid behavior owing to their easy phase tunability by non-thermal parameters, such as magnetic field $B$, external pressure $P$ or chemical substitutions $x$. Heavy-fermion compounds contain a lattice of quasi-particle states formed by hybridization between localized magnetic moments from the $f$ orbitals of heavy elements and the surrounding conduction electrons. The resulting Fermi-liquid state formed below a characteristic temperature $T_{coh}$ gains an effective mass much larger than the free-electron mass \cite{Wirth2016,Coleman2006}. In addition, the competition between the Ruderman-Kittel-Kasuya-Yosida (RKKY)  interaction between local moments and the Kondo interaction between the local moments and itinerant carriers commonly results in a complex phase diagram for these correlated electron systems.

The  compound investigated here, CeRhIn$_5$, belongs to the 1-1-5 heavy-fermion family Ce$M$In$_5$ ($M$ = Co, Rh, In), which have a quasi-2D lattice structure with an alternating layer of CeIn$_3$ and $M$In$_2$ planes stacked along the tetragonal $c$-axis. This arrangement induces strong interaction among the in-plane Ce ions \cite{Kumar2005}. CeRhIn$_5$ undergoes an antiferromagnetic (AFM) transition at $T_N = 3.8$~K \cite{Paglione2008} that is gradually suppressed by moderate external pressure and vanishes around 2~GPa \cite{Yashima2007,Raymond2008}.
Simultaneously a superconducting state appears that reaches its maximum transition temperature $T_c = 2.2$~K at $P \approx 2.5$~GPa. Superconductivity coexists with the AFM order \cite{Knebel2011, Nakajima2007,Knebel2008}
and there are reports that traces even extend down to ambient pressure with $T_c< 0.1$~K \cite{Chen2006,Paglione2008}.
The phase co-existence around the critical point makes CeRhIn$_5$ an interesting heavy-fermion system to study the intriguing relation between the AFM quantum criticality and the unconventional superconductivity
\cite{Donath2007,Mendonca2008,Knebel2011Comptes,Seo2015,Jung2018}.

In addition, non-Fermi liquid signatures are investigated by magnetic, thermodynamic and transport measurements where hydrostatic pressure and Sn doping allows tuning through the transition \cite{Mathur1998,Hegger2000, Shishido2005,Raymond2014}. The substitution of Sn for In corresponds to electron doping and acts as a positive pressure in the phase diagram. Increasing $x$
in CeRh(In$_{1-x}$Sn$_x$)$_5$  gradually suppresses $T_N$ until a critical concentration around $x$ = 7 \%, where the magnetic transition vanishes \cite{Bauer2006,Park2020,Donath2009}.
Superconductivity appears only under the application of external pressure $P$ with a reduced maximum value of $T_c$ \cite{Mendonca2008}. For that reason, the antiferromagnetic quantum critcal point (QCP) can be studied at ambient pressure without being affected by superconductivity.
Since, Sn has an extra $p$ electron compared to In, the density of the state of conduction electrons increases with increasing Sn substitution, consequently enhancing the $c$-$f$ hybridization. This leads to a monotonic suppression of the magnetic order and enhancement of the Kondo interaction.

\begin{figure}[t]
    \centering
    \includegraphics[width=0.45\textwidth]
    {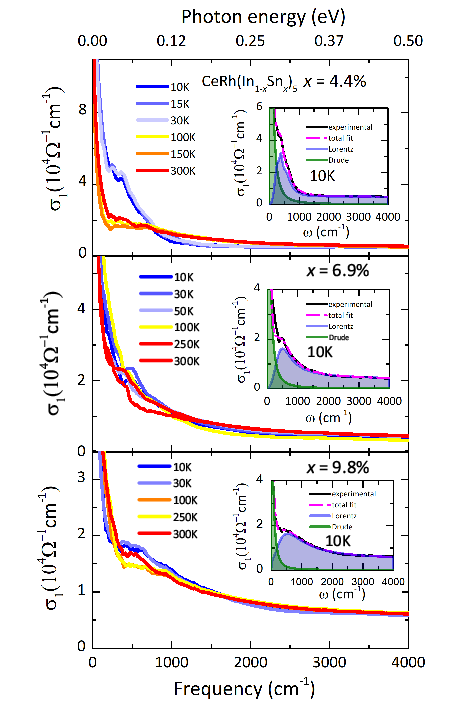}
    \caption{Real part of the optical conductivity of CeRh(Sn$_{1-x}$In$_x$)$_5$ obtained from reflectivity measurements at different temperatures, with Sn concentrations of $x = 4.4\%$, 6.9\%\ and 9.8\%, respectively. Note the different ordinates. The inset demonstrates how the 10~K spectra can be modelled by Drude and Lorentz terms. The MIR feature that characterises the formation of heavy fermions develops below 30~K, which gets more prominently broadened towards 9.8\%\ Sn concentration.}
    \label{opticalconductivity}
\end{figure}

Here we investigate the behavior of the massive quasi-particles in the vicinity of the antiferromagnetic QCP. Besides temperature-dependent transport and magnetic measurements, we utilize infrared optical spectroscopy which is a very powerful tool to study the frequency-dependent response of collective excitations in heavy fermionic materials since the energy corresponds to the hybridization gap.
This way we can probe the excitations across the hybridized bands and the quasiparticle excitations near the Fermi level that occur with the formation of a coherent state \cite{Basov2011,Kimura2012,Chen2016}.
At elevated temperatures, the optical conductivity of heavy fermionic compounds is analogous to that of  simple metals and can typically be described by the Drude model
\begin{equation}
\hat{\sigma}(\omega) = \sigma_1 (\omega)+{\rm i}\sigma_2(\omega) =
\frac{\sigma_{\rm dc}}{1+i\omega \tau}
\quad ,
\end{equation}
which depends on the dc conductivity $\sigma_{\rm dc}$ and the carrier relaxation time $\tau$ \cite{DresselGruner2002}. In the coherent state the electrodynamics is characterized by the emergence of a mid-infrared (MIR) peak that provides direct information on the magnitude of the hybridization strength $V$, which is related to the energy gap corresponding to the Kondo temperature $T_K$ and the width of the conduction band $W$ through $E_{\rm MIR} \simeq 2V \propto \sqrt{T_K W}$. Concomitantly, the Drude peak, representing the intraband response is strongly reduced in width. The narrow Drude peak that forms below the coherent temperature $T_{\rm coh}$, owing to the emerging heavy fermion quasiparticles, has a renormalized heavy effective mass $m^*$ and a long relaxation time $\tau^*$ \cite{DresselGruner2002,Degiorgi1999,Kimura2016,Okamura2007}.
By closely inspecting the $c$-$f$ hybridization in CeRh(In$_{1-x}$Sn$_x$)$_5$, we can reveal the effect of Sn doping on the coherent phenomenon. Using the extended Drude model we determine the frequency-dependent scattering rate and effective mass \cite{DresselGruner2002,Dordevic2001}:
\begin{subequations}
\begin{eqnarray}
\Gamma(\omega) = \frac{1}{\tau^*}=\frac{\omega_p^2}{4\pi} \frac{\sigma_1(\omega)}{|\hat{\sigma}(\omega)|^2} \label{eq:extDrude_scattering}\\
\frac{m^*(\omega)}{m_b} = \frac{\omega_p^2}{4\pi} \frac{\sigma_2(\omega)/\omega}{|\hat{\sigma}(\omega)|^2}
\quad ,
\label{eq:extDrude_mass}
\end{eqnarray}
\label{eq:extDrude}
\end{subequations}
where $\omega_p^2={4\pi Ne^2}/{m_b}$ denotes the plasma frequency with $N$ the carrier density and $m_b$ the band mass. $\omega_p$ is also related to the integration of $ \sigma _{1}(\omega )$ over frequency, the optical spectral weight, via $\int \sigma _{1}(\omega)d\omega =\omega_p ^{2}/8$. Heavy fermions are known as model systems that in general obey Landau's Fermi-liquid theory with a temperature and frequency dependent scattering rate $\Gamma(T,\omega) \propto (2\pi k_BT^2)+(\hbar\omega)^2$; however, deviation can occur in the proximity to the QCP where coupling to itinerant spin fluctuations may cause a week Kondo screening \cite{Scheffler2005,Kimura2006,Scheffler2013,Dressel2002Letter,*Dressel2002,Lee2008}.

In the present work, we explore the quantum critical regime of Sn doped CeRh(In$_{1-x}$Sn$_x$)$_5$. We observe that the antiferromagnetically ordered state in $x=4.4\%$ is suppressed towards $x=9.8\%$, where the heavy fermionic nature strengthens with an enhanced hybridization as inferred from infrared optical conductivity. The fingerprints of QCP influenced non-Fermi liquid nature of these compounds is noticed from temperature dependent resistivity and magnetization data, in addition to an observed anomalous enhancement of NFL temperature regime with increasing Sn concentration.

\section{Experimental details}

Single crystals of CeRh(In$_{1-x}$Sn$_x$)$_5$ were synthesized by the in-flux method, as described elsewhere \cite{Park2020,Bauer2006}. With the help of energy-dispersive
x-ray-spectroscopy the obtained Sn concentration was determined as $x=4.4\%$, 6.9\%\ and  9.8\%.
The infrared reflectivity of polished crystals of approximately $3 \times 3~{\rm mm}^2$ reflecting planes was measured in the frequency range from 80~cm$^{-1}$ to 20,000~cm$^{-1}$ (note: 8065~cm$^{-1}$ corresponds to photon energy of 1~eV)
as a function of temperature from $T=300$~K down to 10~K.
The far-infrared reflectivity measurements were performed using a Bruker IFS 66v spectrometer and a custom-built cryostat. The gold overcoating technique was applied to obtain the absolute value of reflectivity. In addition, we employed a Bruker Vertex 80v attached to a Hyperion IR microscope for
the mid- and near-infrared range $\omega/(2\pi c) > 650~{\rm cm}^{-1}$; here the sample was freshly evaporated by gold as a reference. The optical conductivity is calculated via the Kramers-Kronig analysis with a Hagen-Ruben extrapolation below 80~cm$^{-1}$ and x-ray scattering function for the high-energy region \cite{Tanner2015}.

The magnetic properties of the crystals were measured as a function of temperature in a superconducting quantum interference device (SQUID) magnetometer (M/S Quantum Design) with the external magnetic field  $B = 0.2$~T  perpendicular to the $ab$ plane. The temperature dependence of resistivity was probed by standard four-point technique in a helium-bath cryostat down to 2~K.

\section{Results and Discussion}
\subsection{Optical Properties}
Fig.~\ref{opticalconductivity} displays the temperature-dependent optical conductivity of Sn-doped CeRhIn$_5$ at three concentrations. All compounds possess a highly metallic nature in accord with the notably large dc conductivity values that reach the order of 10$^4 (\Omega {\rm cm})^{-1}$. A weak hybridizing nature of  pristine CeRhIn$_5$ is reported previously \cite{Mena2005,Burch2007,Okamura2015}; the behavior resembles other magnetically-ordered heavy-fermion compounds like CeIn$_3$
due to weak Kondo coupling \cite{Iizuka2012,Chen2016,Dressel2002}.
The hybridization gap in $\sigma_1(\omega)$ forms at significantly lower energy compared to other strongly hybridizing compounds such as CeCoIn$_5$ \cite{Singley2002}.

\begin{figure}[t]
    \centering
    \includegraphics[width=0.45\textwidth]{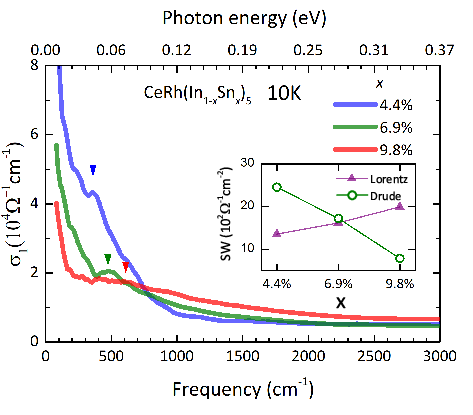}
    \caption{Comparison of the low-temperature conductivity of CeRh(In$_{1-x}$Sn$_x$)$_5$ with $x=4.4\%$, 6.9\%\ and  9.8\%. The arrows illustrate the evolution of $E_{\rm MIR}$ with increasing Sn concentration indicating an enhancement of $c$-$f$ hybridization strength.
    The insert presents the transfer of spectral weight from the Drude to the Lorentz component with increasing Sn concentration.}
    \label{10kopticalconductivity}
\end{figure}

\begin{figure}[t]
    \centering
    \includegraphics[width=0.4\textwidth]{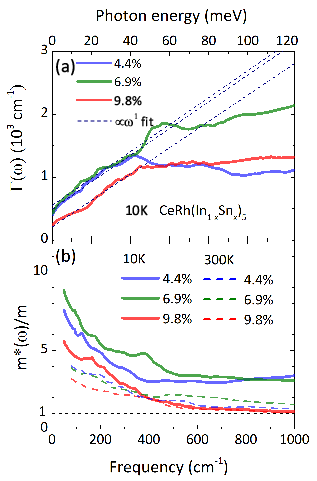}
    \caption{ (a)~The frequency-dependent scattering rate $\Gamma(\omega)$ of  CeRh(In$_{1-x}$Sn$_x$)$_5$ reveals a linear $\omega$ dependence indicating a non-Fermi liquid behavior. The dotted lines represent $\Gamma \propto \omega$ for the three compositions.
    (b)~The corresponding frequency-dependent effective mass $m^*(\omega)$ diverges at low frequencies.}
    \label{scatteringrateandeffectivemass}
\end{figure}

The high-temperature optical response is dominated by the broad Drude behavior of the conduction electrons. With decreasing temperature, a hybridization gap is formed resulting in the emergence of a MIR peak due to excitation across the hybridization gap; the feature is modelled by a Lorentzian absorption as demonstrated by the blue area in the insets of Fig.~\ref{opticalconductivity}.
The excitations close to the Fermi surface due to the heavy-fermion quasiparticles are modelled by a Drude term represented in the green-shaded region of these insets. This feature is observed below 30~K, and becomes more prominent at $T=10$~K.
For the compound with 4.4\%\ of Sn, the hybridization gap opens around 30 K.
The spectra exhibit a distinct character between high and low temperatures;
in addition, the weak peak forming at low frequency around 350~cm$^{-1}$ implies a small hybridization
in CeRh(In$_{0.956}$Sn$_{0.044}$)$_5$. In contrast, the $x=6.9\%$ system shows well-defined features with a MIR band around 450~cm$^{-1}$.  For the largest Sn doping ($x=9.8~\%$) the gap opens most prominently and a broad MIR peak centered around 550~cm$^{-1}$. For $T>$ 30 K no characteristic feature is present in the MIR.

In Fig.~\ref{10kopticalconductivity} the optical conductivity of the three compounds is compared for $T=10$~K. With increasing Sn concentration a clear shift of $E_{\rm MIR}$ to higher energies is observed. When describing the spectra by a Drude and Lorentz terms, as illustrated in the insets of Fig.~\ref{opticalconductivity}, a pronounced redistribution of spectral weight as the Sn doping increases [inset of Fig.~\ref{10kopticalconductivity}].
The Drude contribution becomes narrower because the scattering rate is reduced with larger amount of Sn.  The local moments are better screened by the conduction electrons giving rise to a coherent scattering from heavy-fermion quasiparticles. This is also indicated by a reduced $\omega_p$ in the order 31, 29, and $19 \times 10^3$~cm$^{-1}$, with increasing Sn concentration. For 4.4\%\ of Sn the MIR peak is comparatively narrow, while it broadens for $x= 9.8\%$. Correspondingly, the conduction band gets wider because more $f$-electrons contribute to the itinerant charge carriers. Since $E_{\rm MIR}$ depends on the conduction band width, also the hybridization becomes stronger.
Our results provide evidence that CeRh(In$_{1-x}$Sn$_x$)$_5$ is in an only weakly hybridized state for Sn concentration of $x=4.4\%$ while it becomes strongly hybridized for  9.8\%\ of Sn  substitution.
Here the hybridization is comparable to the pristine CeCoIn$_5$, where the MIR peak occurs 600~cm$^{-1}$ \cite{Okamura2015}. The $x=9.8\%$ compound exhibits a hybridization similar to the one obtained by applying 2~GPa external pressure \cite{Okamura2019}.

The frequency-dependent scattering rate and effective mass were calculated using the extended Drude model \cite{Dressel2002} given in Eqs.~(\ref{eq:extDrude}). The scattering rate $\Gamma(\omega)$ displayed in Fig.~\ref{scatteringrateandeffectivemass}(a) exhibits a pronounced frequency dependence at low energies which are caused by many-body effects due to the $c$-$f$ hybridization.
For all compounds, the scattering rate increases with frequency, shows a hump,
before it basically saturates at high frequencies. This behavior in the scattering spectrum is caused by transitions across the hybridisation gap, reducing the effect of scattering.
The hump is interpreted as the gap in the optical conductivity \cite{Kimura2006}.
The large scattering rate at high frequencies is related to scattering of itinerant electrons by the Ce 4$f$ localized moments. The maxima in the scattering rate shift to higher energies with doping.
These maxima correspond to the frequency where the enhancement of the effective mass $m^*(\omega)$ starts
[Fig.~\ref{scatteringrateandeffectivemass}(b)].
Interestingly, the scattering rate of the compound with 9.8\%\ Sn doping is the lowest.
At $T=10$~K the compounds are in a coherent regime and the low-frequency scattering
depends on electron scattering from spin fluctuations rather than scattering from ions.
At low frequencies the scattering rate follows an almost linear frequency dependence $\Gamma(\omega) \propto \omega$ indicating a  non-Fermi liquid behavior for all three compounds at $T=10$~K.
Also, the effective mass is strongly enhanced at low temperatures. Fig.~\ref{scatteringrateandeffectivemass}(b)
shows a pronounced low-frequency divergence that starts around 200~cm$^{-1}$.
The behavior of $m^*(\omega)/m$ does not saturate for $\omega \rightarrow 0$,
which might indicate scattering from critical fluctuations.

\subsection{Transport and Magnetic Properties}
\begin{figure}[t]
    \centering
    \includegraphics[width=0.45\textwidth]{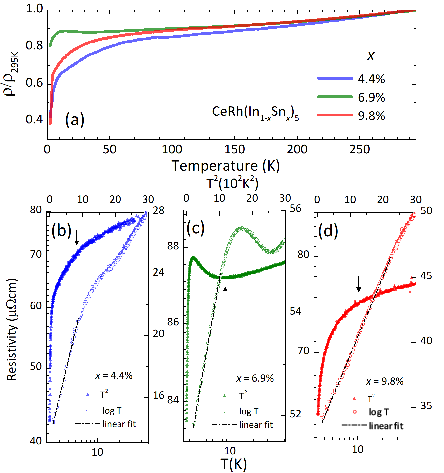}
    \caption{(a) Normalized resistivity for three compounds indicating a drop in resistivity with the formation of Kondo lattice and a sharp drop around the coherent temperature $T_{\rm coh}$. (b-d)~ Low-temperature resistivity of CeRh(In$_{1-x}$Sn$_x$)$_5$. The left and lower axes use logarithmic scales. The upper and right axes correspond to a $\rho(T)$ {\it vs.} $T^2$ representation. The arrows indicate the lower-temperature limit of the Fermi-liquid behavior, where $\rho(T)\propto T^2$ is observed.}
    \label{resistivity}
\end{figure}

\begin{figure}[t]
    \centering
    \includegraphics[width=0.45\textwidth]{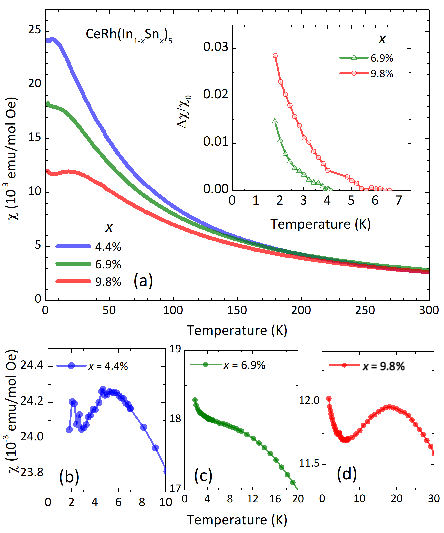}
    \caption{ (a)~ Magnetic susceptibility as a function of temperature for different Sn concentrations in CeRh(In$_{1-x}$Sn$_x$)$_5$.
    With increasing $x$ the susceptibility drops in magnitude and a saturation is observed as heavy-fermion quasiparticles form.
    The inset shows a rise in susceptibility for Sn 6.9 $\%$ and Sn 9.8 $\%$. (b-d) Temperature-dependent magnetization of CeRh(In$_{1-x}$Sn$_x$)$_5$ with $x=4.4\%$, 6.9\%\ and  9.8\%. No clear saturation is found at low temperatures.
    For 4.4\%\ of Sn an antiferromagnetic transition is seen at 2~K.
    An anomaly-like hump broadens and shifts to higher temperature with increasing Sn concentration.}
    \label{susceptability}
\end{figure}

Further information of the non-Fermi-liquid behavior can be obtained from resistivity and magnetization measurements that reveal the detailed temperature-dependent properties near the QCP.
For all three compounds, the resistivity $\rho(T)$ presented in Fig.~\ref{resistivity}(a)
drops sharply below the coherence temperature $T_{\rm coh}$, because of coherent scattering due to the formation of a Kondo lattice. The coherence temperature is found to be around 35~K for $x=6.9\%$ and 9.8\%, while it is around 14~K for 4.4\%\ of Sn. The low-temperature resistivity is plotted in Fig.~\ref{resistivity}~(b-d) using log-log scales in order to examine power-law behaviors.
For the three Sn compositions, 4.4\%, 6.9\%\ and 9.8\%, a power law $\rho(T)\propto T^n$ is found up to 5.5~K, 7.5~K and 25~K with exponents $n = 0.4$, 0.058 and 0.12, respectively. Such a sub - $T$ linear dependence is reported previously \cite{Seo2015}. It is interesting that below the crossover temperature $T^*$, $\rho(T)$ follows a single power law down to the lowest measured temperature. This deviation from Landau's $T^2$ behavior provides strong evidence, that
these compounds are close to a QCP.
Above these temperature, the materials obey a Fermi-liquid behavior, above the region indicated by arrows in the $T^2$-presentation of the upper and right axes in Fig.~\ref{resistivity}(b-d).

The temperature dependence of the magnetic susceptibility in the temperature range from $T = 2$ to 300 K
is displayed in Fig.~\ref{susceptability}. At high temperature, the susceptibility follows a Curie behavior due to local moments originating from Ce ions.
Details of the high-temperature behavior have been extensively discussed in earlier reports \cite{Park2020}. The saturation of the susceptibility reflecting the Pauli paramagnetism evidences Fermi-liquid behavior caused by the Kondo screening effect, where the localised magnetic moments are screened by the itinerant electrons. The susceptibility of the three samples nearly merge at high temperatures, but the curves significantly deviate from each other at low temperatures.
The magnitude of $\chi$ monotonically decreases with increasing Sn substitution
indicating that the increase in $c$-$f$ hybridization favours the Kondo screening of the local moments, and hence results into a reduced magnetization at low temperatures.

In the following, we will focus on the temperature dependence of the susceptibility at low temperatures,
which is plotted in Fig.~\ref{susceptability} (b-d). In case of  $x = 4.4\%$ the susceptibility slowly increases
upon cooling down to $T \sim 5$~K, and then it starts to decrease showing a shallow maximum.
Around $T = 2.1$~K, $\chi(T)$ exhibits a sharp peak that corresponds to the AFM transition, confirming earlier reports \cite{Park2020}. For $x = 6.9\%$ the susceptibility tends to saturate and finally shows a weak divergence at very low temperatures; the shallow maximum occurs around $T^* \sim 10$~K.
For the higher Sn-doping under inspection here, $x = 9.8\%$, the susceptibility goes through a broad maximum at $T^* \sim 18$~K and a minimum around 8 K before it diverges rapidly at low temperatures.
In the inset of Fig.~\ref{susceptability}(a) the divergence of $\chi(T)$ is compared for $x = 6.9\%$ and 9.8\%\ for  $T < 8$~K. Here, we plot the susceptibility increase $\Delta \chi(T)$ = $\chi$(T) - $\chi$(0) as a function of temperature where $\chi$(0) is the value at the temperature just above the rise.
Although we cannot present a definite explanation, we note that such anomalous temperature dependence of the susceptibility has been observed due to Kondo screening of the magnetic moments below the Kondo temperature \cite{Luo2014}. On the other hand, a similar temperature dependence of $\chi(T)$ was seen in YbRhSi$_2$
where the evolution of quantum criticality was investigated upon chemical pressure \cite{Friedemann2009}. Signatures of a Kondo breakdown were observed within the magnetically ordered state.
When the Kondo effect vanishes the $f$-states become localized near the QCP, resulting in a Fermi surface reconstruction. In the parent compound YbRhSi$_2$, the Kondo beakdown is believed to occur at the QCP; and chemical pressure in terms of doping induces a separation of Kondo breakdown and the QCP.
Previous experiments \cite{Bauer2006} have observed a similar weak divergence of the spin susceptibility in Sn-doped CeRhIn$_5$ and related it to the localized/itinerant crossover. Such anomalous behavior needs further attention covering even lower temperatures.


\section{Conclussions}

We have performed infrared optical experiments, electrical resistivity and magnetic susceptibility measurements to probe the hybridization gap and examined the non-Fermi-liquid properties of the heavy fermionic compound  CeRh(In$_{1-x}$Sn$_x$)$_5$ with different concentrations of Sn: $x=4.4\%$, 6.9\%\ and 9.8\%. The build-up of a mid-infrared peak in the optical conductivity observed in all compounds below $T=30$~K reveals the formation of a hybridization gap. The energy related to the gaps shifts to higher frequencies with doping
revealing a clear enhancement of the hybridization strength. At low temperatures ($T=10$~K) all three compounds exhibit a frequency-dependent scattering rate $\Gamma(\omega)$ with a linear increase up to approximately 400~cm$^{-1}$, indicating increasing spin fluctuations upon approaching the quantum critical point. Concomitantly the frequency-dependent effective mass $m^*(\omega)$ does not saturate for lowering the frequency. The magnetic susceptibility reveals antiferromagnetic order below 2~K for  4.4\%\ Sn. At higher concentrations, $x = 6.9\%$ and 9.8\%, $\chi(T)$ increases below $T=3$~K and 6 K, respectively.
This effect may be linked to the proximity to the quantum critical point.
We also see an anomalous hump in $\chi(T)$ that broadens and moves to higher temperatures upon doping.
The low-temperature resistivity behaves in a non-Fermi-liquid manner; this effect is present up to 25~K for the compound with the highest Sn concentration.

\acknowledgments

The authors acknowledge the technical supports from Gabriele Untereiner.
We thank Pascal Puphal for discussion and additional characterization.
Run Yang appreciates a fellowship by the Alexander von Humboldt Foundation (Germany).


\begin{thebibliography}{51}%
\makeatletter
\providecommand \@ifxundefined [1]{%
 \@ifx{#1\undefined}
}%
\providecommand \@ifnum [1]{%
 \ifnum #1\expandafter \@firstoftwo
 \else \expandafter \@secondoftwo
 \fi
}%
\providecommand \@ifx [1]{%
 \ifx #1\expandafter \@firstoftwo
 \else \expandafter \@secondoftwo
 \fi
}%
\providecommand \natexlab [1]{#1}%
\providecommand \enquote  [1]{``#1''}%
\providecommand \bibnamefont  [1]{#1}%
\providecommand \bibfnamefont [1]{#1}%
\providecommand \citenamefont [1]{#1}%
\providecommand \href@noop [0]{\@secondoftwo}%
\providecommand \href [0]{\begingroup \@sanitize@url \@href}%
\providecommand \@href[1]{\@@startlink{#1}\@@href}%
\providecommand \@@href[1]{\endgroup#1\@@endlink}%
\providecommand \@sanitize@url [0]{\catcode `\\12\catcode `\$12\catcode
  `\&12\catcode `\#12\catcode `\^12\catcode `\_12\catcode `\%12\relax}%
\providecommand \@@startlink[1]{}%
\providecommand \@@endlink[0]{}%
\providecommand \url  [0]{\begingroup\@sanitize@url \@url }%
\providecommand \@url [1]{\endgroup\@href {#1}{\urlprefix }}%
\providecommand \urlprefix  [0]{URL }%
\providecommand \Eprint [0]{\href }%
\providecommand \doibase [0]{https://doi.org/}%
\providecommand \selectlanguage [0]{\@gobble}%
\providecommand \bibinfo  [0]{\@secondoftwo}%
\providecommand \bibfield  [0]{\@secondoftwo}%
\providecommand \translation [1]{[#1]}%
\providecommand \BibitemOpen [0]{}%
\providecommand \bibitemStop [0]{}%
\providecommand \bibitemNoStop [0]{.\EOS\space}%
\providecommand \EOS [0]{\spacefactor3000\relax}%
\providecommand \BibitemShut  [1]{\csname bibitem#1\endcsname}%
\let\auto@bib@innerbib\@empty
\bibitem [{\citenamefont {Si}\ and\ \citenamefont {Steglich}(2010)}]{Si2010}%
  \BibitemOpen
  \bibfield  {author} {\bibinfo {author} {\bibfnamefont {Q.}~\bibnamefont
  {Si}}\ and\ \bibinfo {author} {\bibfnamefont {F.}~\bibnamefont {Steglich}},\
  }\bibfield  {title} {\bibinfo {title} {Heavy fermions and quantum phase
  transitions},\ }\href {https://doi.org/10.1126/science.1191195} {\bibfield
  {journal} {\bibinfo  {journal} {Science}\ }\textbf {\bibinfo {volume}
  {329}},\ \bibinfo {pages} {1161} (\bibinfo {year} {2010})}\BibitemShut
  {NoStop}%
\bibitem [{\citenamefont {Paschen}\ and\ \citenamefont
  {Si}(2021)}]{Paschen2021}%
  \BibitemOpen
  \bibfield  {author} {\bibinfo {author} {\bibfnamefont {S.}~\bibnamefont
  {Paschen}}\ and\ \bibinfo {author} {\bibfnamefont {Q.}~\bibnamefont {Si}},\
  }\bibfield  {title} {\bibinfo {title} {Quantum phases driven by strong
  correlations},\ }\href {https://doi.org/10.1038/s42254-020-00262-6}
  {\bibfield  {journal} {\bibinfo  {journal} {Nat. Rev. Phys.}\ }\textbf
  {\bibinfo {volume} {3}},\ \bibinfo {pages} {9} (\bibinfo {year}
  {2021})}\BibitemShut {NoStop}%
\bibitem [{\citenamefont {Dressel}(2011)}]{Dressel2011}%
  \BibitemOpen
  \bibfield  {author} {\bibinfo {author} {\bibfnamefont {M.}~\bibnamefont
  {Dressel}},\ }\bibfield  {title} {\bibinfo {title} {Quantum criticality in
  organic conductors? {{Fermi}} liquid versus {{non-Fermi-liquid}} behaviour},\
  }\href {https://doi.org/10.1088/0953-8984/23/29/293201} {\bibfield  {journal}
  {\bibinfo  {journal} {J. Phys.: Condens. Matter}\ }\textbf {\bibinfo {volume}
  {23}},\ \bibinfo {pages} {293201} (\bibinfo {year} {2011})}\BibitemShut
  {NoStop}%
\bibitem [{\citenamefont {Dressel}\ and\ \citenamefont
  {Tomić}(2020)}]{Dressel2020}%
  \BibitemOpen
  \bibfield  {author} {\bibinfo {author} {\bibfnamefont {M.}~\bibnamefont
  {Dressel}}\ and\ \bibinfo {author} {\bibfnamefont {S.}~\bibnamefont
  {Tomić}},\ }\bibfield  {title} {\bibinfo {title} {Molecular quantum
  materials: electronic phases and charge dynamics in two-dimensional organic
  solids},\ }\href {https://doi.org/10.1080/00018732.2020.1837833} {\bibfield
  {journal} {\bibinfo  {journal} {Adv. Phys.}\ }\textbf {\bibinfo {volume}
  {69}},\ \bibinfo {pages} {1} (\bibinfo {year} {2020})}\BibitemShut {NoStop}%
\bibitem [{\citenamefont {L\"ohneysen}\ \emph {et~al.}(2007)\citenamefont
  {L\"ohneysen}, \citenamefont {Rosch}, \citenamefont {Vojta},\ and\
  \citenamefont {W\"olfle}}]{Lohneysen2007}%
  \BibitemOpen
  \bibfield  {author} {\bibinfo {author} {\bibfnamefont {H.~v.}\ \bibnamefont
  {L\"ohneysen}}, \bibinfo {author} {\bibfnamefont {A.}~\bibnamefont {Rosch}},
  \bibinfo {author} {\bibfnamefont {M.}~\bibnamefont {Vojta}},\ and\ \bibinfo
  {author} {\bibfnamefont {P.}~\bibnamefont {W\"olfle}},\ }\bibfield  {title}
  {\bibinfo {title} {Fermi-liquid instabilities at magnetic quantum phase
  transitions},\ }\href {https://doi.org/10.1103/RevModPhys.79.1015} {\bibfield
   {journal} {\bibinfo  {journal} {Rev. Mod. Phys.}\ }\textbf {\bibinfo
  {volume} {79}},\ \bibinfo {pages} {1015} (\bibinfo {year}
  {2007})}\BibitemShut {NoStop}%
\bibitem [{\citenamefont {Kuga}\ \emph {et~al.}(2018)\citenamefont {Kuga},
  \citenamefont {Matsumoto}, \citenamefont {Okawa}, \citenamefont {Suzuki},
  \citenamefont {Tomita}, \citenamefont {Sone}, \citenamefont {Shimura},
  \citenamefont {Sakakibara}, \citenamefont {Nishio-Hamane}, \citenamefont
  {Karaki}, \citenamefont {Takata}, \citenamefont {Matsunami}, \citenamefont
  {Eguchi}, \citenamefont {Taguchi}, \citenamefont {Chainani}, \citenamefont
  {Shin}, \citenamefont {Tamasaku}, \citenamefont {Nishino}, \citenamefont
  {Yabashi}, \citenamefont {Ishikawa},\ and\ \citenamefont
  {Nakatsuji}}]{Kuga2018}%
  \BibitemOpen
  \bibfield  {author} {\bibinfo {author} {\bibfnamefont {K.}~\bibnamefont
  {Kuga}}, \bibinfo {author} {\bibfnamefont {Y.}~\bibnamefont {Matsumoto}},
  \bibinfo {author} {\bibfnamefont {M.}~\bibnamefont {Okawa}}, \bibinfo
  {author} {\bibfnamefont {S.}~\bibnamefont {Suzuki}}, \bibinfo {author}
  {\bibfnamefont {T.}~\bibnamefont {Tomita}}, \bibinfo {author} {\bibfnamefont
  {K.}~\bibnamefont {Sone}}, \bibinfo {author} {\bibfnamefont {Y.}~\bibnamefont
  {Shimura}}, \bibinfo {author} {\bibfnamefont {T.}~\bibnamefont {Sakakibara}},
  \bibinfo {author} {\bibfnamefont {D.}~\bibnamefont {Nishio-Hamane}}, \bibinfo
  {author} {\bibfnamefont {Y.}~\bibnamefont {Karaki}}, \bibinfo {author}
  {\bibfnamefont {Y.}~\bibnamefont {Takata}}, \bibinfo {author} {\bibfnamefont
  {M.}~\bibnamefont {Matsunami}}, \bibinfo {author} {\bibfnamefont
  {R.}~\bibnamefont {Eguchi}}, \bibinfo {author} {\bibfnamefont
  {M.}~\bibnamefont {Taguchi}}, \bibinfo {author} {\bibfnamefont
  {A.}~\bibnamefont {Chainani}}, \bibinfo {author} {\bibfnamefont
  {S.}~\bibnamefont {Shin}}, \bibinfo {author} {\bibfnamefont {K.}~\bibnamefont
  {Tamasaku}}, \bibinfo {author} {\bibfnamefont {Y.}~\bibnamefont {Nishino}},
  \bibinfo {author} {\bibfnamefont {M.}~\bibnamefont {Yabashi}}, \bibinfo
  {author} {\bibfnamefont {T.}~\bibnamefont {Ishikawa}},\ and\ \bibinfo
  {author} {\bibfnamefont {S.}~\bibnamefont {Nakatsuji}},\ }\bibfield  {title}
  {\bibinfo {title} {Quantum valence criticality in a correlated metal},\
  }\href {https://doi.org/10.1126/sciadv.aao3547} {\bibfield  {journal}
  {\bibinfo  {journal} {Sci. Adv.}\ }\textbf {\bibinfo {volume} {4}},\ \bibinfo
  {pages} {eaao3547} (\bibinfo {year} {2018})}\BibitemShut {NoStop}%
\bibitem [{\citenamefont {Wirth}\ and\ \citenamefont
  {Steglich}(2016)}]{Wirth2016}%
  \BibitemOpen
  \bibfield  {author} {\bibinfo {author} {\bibfnamefont {S.}~\bibnamefont
  {Wirth}}\ and\ \bibinfo {author} {\bibfnamefont {F.}~\bibnamefont
  {Steglich}},\ }\bibfield  {title} {\bibinfo {title} {Exploring heavy fermions
  from macroscopic to microscopic length scales},\ }\href
  {https://doi.org/10.1038/natrevmats.2016.66} {\bibfield  {journal} {\bibinfo
  {journal} {Nat. Rev. Mater.}\ }\textbf {\bibinfo {volume} {1}},\ \bibinfo
  {pages} {1} (\bibinfo {year} {2016})}\BibitemShut {NoStop}%
\bibitem [{\citenamefont {Coleman}(2006)}]{Coleman2006}%
  \BibitemOpen
  \bibfield  {author} {\bibinfo {author} {\bibfnamefont {P.}~\bibnamefont
  {Coleman}},\ }\bibfield  {title} {\bibinfo {title} {Heavy fermions: Electrons
  at the edge of magnetism},\ }\href {https://arxiv.org/abs/cond-mat/0612006}
  {\bibfield  {journal} {\bibinfo  {journal} {arXiv:0612006}\ } (\bibinfo
  {year} {2006})}\BibitemShut {NoStop}%
\bibitem [{\citenamefont {Kumar}\ \emph {et~al.}(2005)\citenamefont {Kumar},
  \citenamefont {Kohlmann}, \citenamefont {Light}, \citenamefont {Cornelius},
  \citenamefont {Raghavan}, \citenamefont {Darling},\ and\ \citenamefont
  {Sarrao}}]{Kumar2005}%
  \BibitemOpen
  \bibfield  {author} {\bibinfo {author} {\bibfnamefont {R.}~\bibnamefont
  {Kumar}}, \bibinfo {author} {\bibfnamefont {H.}~\bibnamefont {Kohlmann}},
  \bibinfo {author} {\bibfnamefont {B.}~\bibnamefont {Light}}, \bibinfo
  {author} {\bibfnamefont {A.}~\bibnamefont {Cornelius}}, \bibinfo {author}
  {\bibfnamefont {V.}~\bibnamefont {Raghavan}}, \bibinfo {author}
  {\bibfnamefont {T.}~\bibnamefont {Darling}},\ and\ \bibinfo {author}
  {\bibfnamefont {J.}~\bibnamefont {Sarrao}},\ }\bibfield  {title} {\bibinfo
  {title} {The crystal structure of {{CeRhIn$_5$}} under pressure},\ }\href
  {https://doi.org/10.1016/j.physb.2005.01.081} {\bibfield  {journal} {\bibinfo
   {journal} {Physica B}\ }\textbf {\bibinfo {volume} {359}},\ \bibinfo {pages}
  {407} (\bibinfo {year} {2005})}\BibitemShut {NoStop}%
\bibitem [{\citenamefont {Paglione}\ \emph {et~al.}(2008)\citenamefont
  {Paglione}, \citenamefont {Ho}, \citenamefont {Maple}, \citenamefont
  {Tanatar}, \citenamefont {Taillefer}, \citenamefont {Lee},\ and\
  \citenamefont {Petrovic}}]{Paglione2008}%
  \BibitemOpen
  \bibfield  {author} {\bibinfo {author} {\bibfnamefont {J.}~\bibnamefont
  {Paglione}}, \bibinfo {author} {\bibfnamefont {P.-C.}\ \bibnamefont {Ho}},
  \bibinfo {author} {\bibfnamefont {M.~B.}\ \bibnamefont {Maple}}, \bibinfo
  {author} {\bibfnamefont {M.~A.}\ \bibnamefont {Tanatar}}, \bibinfo {author}
  {\bibfnamefont {L.}~\bibnamefont {Taillefer}}, \bibinfo {author}
  {\bibfnamefont {Y.}~\bibnamefont {Lee}},\ and\ \bibinfo {author}
  {\bibfnamefont {C.}~\bibnamefont {Petrovic}},\ }\bibfield  {title} {\bibinfo
  {title} {Ambient-pressure bulk superconductivity deep in the magnetic state
  of $\mathrm{Ce}\mathrm{Rh}{\mathrm{in}}_{5}$},\ }\href
  {https://doi.org/10.1103/PhysRevB.77.100505} {\bibfield  {journal} {\bibinfo
  {journal} {Phys. Rev. B}\ }\textbf {\bibinfo {volume} {77}},\ \bibinfo
  {pages} {100505(R)} (\bibinfo {year} {2008})}\BibitemShut {NoStop}%
\bibitem [{\citenamefont {Yashima}\ \emph {et~al.}(2007)\citenamefont
  {Yashima}, \citenamefont {Kawasaki}, \citenamefont {Mukuda}, \citenamefont
  {Kitaoka}, \citenamefont {Shishido}, \citenamefont {Settai},\ and\
  \citenamefont {\ifmmode~\bar{O}\else \={O}\fi{}nuki}}]{Yashima2007}%
  \BibitemOpen
  \bibfield  {author} {\bibinfo {author} {\bibfnamefont {M.}~\bibnamefont
  {Yashima}}, \bibinfo {author} {\bibfnamefont {S.}~\bibnamefont {Kawasaki}},
  \bibinfo {author} {\bibfnamefont {H.}~\bibnamefont {Mukuda}}, \bibinfo
  {author} {\bibfnamefont {Y.}~\bibnamefont {Kitaoka}}, \bibinfo {author}
  {\bibfnamefont {H.}~\bibnamefont {Shishido}}, \bibinfo {author}
  {\bibfnamefont {R.}~\bibnamefont {Settai}},\ and\ \bibinfo {author}
  {\bibfnamefont {Y.}~\bibnamefont {\ifmmode~\bar{O}\else \={O}\fi{}nuki}},\
  }\bibfield  {title} {\bibinfo {title} {Quantum phase diagram of
  antiferromagnetism and superconductivity with a tetracritical point in
  {{CeRhIn}}$_{5}$ in zero magnetic field},\ }\href
  {https://doi.org/10.1103/PhysRevB.76.020509} {\bibfield  {journal} {\bibinfo
  {journal} {Phys. Rev. B}\ }\textbf {\bibinfo {volume} {76}},\ \bibinfo
  {pages} {020509(R)} (\bibinfo {year} {2007})}\BibitemShut {NoStop}%
\bibitem [{\citenamefont {Raymond}\ \emph {et~al.}(2008)\citenamefont
  {Raymond}, \citenamefont {Knebel}, \citenamefont {Aoki},\ and\ \citenamefont
  {Flouquet}}]{Raymond2008}%
  \BibitemOpen
  \bibfield  {author} {\bibinfo {author} {\bibfnamefont {S.}~\bibnamefont
  {Raymond}}, \bibinfo {author} {\bibfnamefont {G.}~\bibnamefont {Knebel}},
  \bibinfo {author} {\bibfnamefont {D.}~\bibnamefont {Aoki}},\ and\ \bibinfo
  {author} {\bibfnamefont {J.}~\bibnamefont {Flouquet}},\ }\bibfield  {title}
  {\bibinfo {title} {Pressure dependence of the magnetic ordering in
  {{CeRhIn$_{5}$}}},\ }\href {https://doi.org/10.1103/PhysRevB.77.172502}
  {\bibfield  {journal} {\bibinfo  {journal} {Phys. Rev. B}\ }\textbf {\bibinfo
  {volume} {77}},\ \bibinfo {pages} {172502} (\bibinfo {year}
  {2008})}\BibitemShut {NoStop}%
\bibitem [{\citenamefont {Knebel}\ \emph
  {et~al.}(2011{\natexlab{a}})\citenamefont {Knebel}, \citenamefont {Buhot},
  \citenamefont {Aoki}, \citenamefont {Lapertot}, \citenamefont {Raymond},
  \citenamefont {Ressouche},\ and\ \citenamefont {Flouquet}}]{Knebel2011}%
  \BibitemOpen
  \bibfield  {author} {\bibinfo {author} {\bibfnamefont {G.}~\bibnamefont
  {Knebel}}, \bibinfo {author} {\bibfnamefont {J.}~\bibnamefont {Buhot}},
  \bibinfo {author} {\bibfnamefont {D.}~\bibnamefont {Aoki}}, \bibinfo {author}
  {\bibfnamefont {G.}~\bibnamefont {Lapertot}}, \bibinfo {author}
  {\bibfnamefont {S.}~\bibnamefont {Raymond}}, \bibinfo {author} {\bibfnamefont
  {E.}~\bibnamefont {Ressouche}},\ and\ \bibinfo {author} {\bibfnamefont
  {J.}~\bibnamefont {Flouquet}},\ }\bibfield  {title} {\bibinfo {title}
  {Antiferromagnetism and superconductivity in {{CeRhIn$_5$}}},\ }\href
  {https://doi.org/10.1143/JPSJS.80SA.SA001} {\bibfield  {journal} {\bibinfo
  {journal} {J. Phys. Soc. Jpn.}\ }\textbf {\bibinfo {volume} {80}},\ \bibinfo
  {pages} {SA001} (\bibinfo {year} {2011}{\natexlab{a}})}\BibitemShut {NoStop}%
\bibitem [{\citenamefont {Nakajima}\ \emph {et~al.}(2007)\citenamefont
  {Nakajima}, \citenamefont {Shishido}, \citenamefont {Nakai}, \citenamefont
  {Shibauchi}, \citenamefont {Behnia}, \citenamefont {Izawa}, \citenamefont
  {Hedo}, \citenamefont {Uwatoko}, \citenamefont {Matsumoto}, \citenamefont
  {Settai}, \citenamefont {Ōnuki}, \citenamefont {Kontani},\ and\
  \citenamefont {Matsuda}}]{Nakajima2007}%
  \BibitemOpen
  \bibfield  {author} {\bibinfo {author} {\bibfnamefont {Y.}~\bibnamefont
  {Nakajima}}, \bibinfo {author} {\bibfnamefont {H.}~\bibnamefont {Shishido}},
  \bibinfo {author} {\bibfnamefont {H.}~\bibnamefont {Nakai}}, \bibinfo
  {author} {\bibfnamefont {T.}~\bibnamefont {Shibauchi}}, \bibinfo {author}
  {\bibfnamefont {K.}~\bibnamefont {Behnia}}, \bibinfo {author} {\bibfnamefont
  {K.}~\bibnamefont {Izawa}}, \bibinfo {author} {\bibfnamefont
  {M.}~\bibnamefont {Hedo}}, \bibinfo {author} {\bibfnamefont {Y.}~\bibnamefont
  {Uwatoko}}, \bibinfo {author} {\bibfnamefont {T.}~\bibnamefont {Matsumoto}},
  \bibinfo {author} {\bibfnamefont {R.}~\bibnamefont {Settai}}, \bibinfo
  {author} {\bibfnamefont {Y.}~\bibnamefont {Ōnuki}}, \bibinfo {author}
  {\bibfnamefont {H.}~\bibnamefont {Kontani}},\ and\ \bibinfo {author}
  {\bibfnamefont {Y.}~\bibnamefont {Matsuda}},\ }\bibfield  {title} {\bibinfo
  {title} {Non-{Fermi} liquid behavior in the magnetotransport of
  {{Ce$M$In$_5$}} ({{$M$}}: {{Co}} and {{Rh}}): striking similarity between
  quasi two-dimensional heavy fermion and high-${{T_c}}$ cuprates},\ }\href
  {https://doi.org/10.1143/JPSJ.76.024703} {\bibfield  {journal} {\bibinfo
  {journal} {J. Phys. Soc. Jpn.}\ }\textbf {\bibinfo {volume} {76}},\ \bibinfo
  {pages} {024703} (\bibinfo {year} {2007})}\BibitemShut {NoStop}%
\bibitem [{\citenamefont {Knebel}\ \emph {et~al.}(2008)\citenamefont {Knebel},
  \citenamefont {Aoki}, \citenamefont {Brison},\ and\ \citenamefont
  {Flouquet}}]{Knebel2008}%
  \BibitemOpen
  \bibfield  {author} {\bibinfo {author} {\bibfnamefont {G.}~\bibnamefont
  {Knebel}}, \bibinfo {author} {\bibfnamefont {D.}~\bibnamefont {Aoki}},
  \bibinfo {author} {\bibfnamefont {J.~P.}\ \bibnamefont {Brison}},\ and\
  \bibinfo {author} {\bibfnamefont {J.}~\bibnamefont {Flouquet}},\ }\bibfield
  {title} {\bibinfo {title} {The quantum critical point in {{CeRhIn$_5$}}: a
  resistivity study},\ }\href {https://doi.org/10.1143/JPSJ.77.114704}
  {\bibfield  {journal} {\bibinfo  {journal} {J. Phys. Soc. Jpn.}\ }\textbf
  {\bibinfo {volume} {77}},\ \bibinfo {pages} {114704} (\bibinfo {year}
  {2008})}\BibitemShut {NoStop}%
\bibitem [{\citenamefont {Chen}\ \emph {et~al.}(2006)\citenamefont {Chen},
  \citenamefont {Matsubayashi}, \citenamefont {Ban}, \citenamefont {Deguchi},\
  and\ \citenamefont {Sato}}]{Chen2006}%
  \BibitemOpen
  \bibfield  {author} {\bibinfo {author} {\bibfnamefont {G.~F.}\ \bibnamefont
  {Chen}}, \bibinfo {author} {\bibfnamefont {K.}~\bibnamefont {Matsubayashi}},
  \bibinfo {author} {\bibfnamefont {S.}~\bibnamefont {Ban}}, \bibinfo {author}
  {\bibfnamefont {K.}~\bibnamefont {Deguchi}},\ and\ \bibinfo {author}
  {\bibfnamefont {N.~K.}\ \bibnamefont {Sato}},\ }\bibfield  {title} {\bibinfo
  {title} {Competitive coexistence of superconductivity with antiferromagnetism
  in {{CeRhIn$_{5}$}}},\ }\href {https://doi.org/10.1103/PhysRevLett.97.017005}
  {\bibfield  {journal} {\bibinfo  {journal} {Phys. Rev. Lett.}\ }\textbf
  {\bibinfo {volume} {97}},\ \bibinfo {pages} {017005(R)} (\bibinfo {year}
  {2006})}\BibitemShut {NoStop}%
\bibitem [{\citenamefont {Donath}\ \emph {et~al.}(2007)\citenamefont {Donath},
  \citenamefont {Gegenwart}, \citenamefont {Steglich}, \citenamefont {Bauer},\
  and\ \citenamefont {Sarrao}}]{Donath2007}%
  \BibitemOpen
  \bibfield  {author} {\bibinfo {author} {\bibfnamefont {J.~G.}\ \bibnamefont
  {Donath}}, \bibinfo {author} {\bibfnamefont {P.}~\bibnamefont {Gegenwart}},
  \bibinfo {author} {\bibfnamefont {F.}~\bibnamefont {Steglich}}, \bibinfo
  {author} {\bibfnamefont {E.~D.}\ \bibnamefont {Bauer}},\ and\ \bibinfo
  {author} {\bibfnamefont {J.~L.}\ \bibnamefont {Sarrao}},\ }\bibfield  {title}
  {\bibinfo {title} {Pressure effect on antiferromagnetism in
  {{CeRhIn$_{5-x}$Sn$_x$}} studied by thermal expansion},\ }\href
  {https://doi.org/10.1016/j.physc.2007.03.030} {\bibfield  {journal} {\bibinfo
   {journal} {Physica C}\ }\textbf {\bibinfo {volume} {460}},\ \bibinfo {pages}
  {661} (\bibinfo {year} {2007})}\BibitemShut {NoStop}%
\bibitem [{\citenamefont {Mendon{\c{c}}a~Ferreira}\ \emph
  {et~al.}(2008)\citenamefont {Mendon{\c{c}}a~Ferreira}, \citenamefont {Park},
  \citenamefont {Sidorov}, \citenamefont {Nicklas}, \citenamefont {Bittar},
  \citenamefont {Lora-Serrano}, \citenamefont {Hering}, \citenamefont {Ramos},
  \citenamefont {Fontes}, \citenamefont {Baggio-Saitovich}, \citenamefont
  {Lee}, \citenamefont {Sarrao}, \citenamefont {Thompson},\ and\ \citenamefont
  {Pagliuso}}]{Mendonca2008}%
  \BibitemOpen
  \bibfield  {author} {\bibinfo {author} {\bibfnamefont {L.}~\bibnamefont
  {Mendon{\c{c}}a~Ferreira}}, \bibinfo {author} {\bibfnamefont
  {T.}~\bibnamefont {Park}}, \bibinfo {author} {\bibfnamefont {V.}~\bibnamefont
  {Sidorov}}, \bibinfo {author} {\bibfnamefont {M.}~\bibnamefont {Nicklas}},
  \bibinfo {author} {\bibfnamefont {E.~M.}\ \bibnamefont {Bittar}}, \bibinfo
  {author} {\bibfnamefont {R.}~\bibnamefont {Lora-Serrano}}, \bibinfo {author}
  {\bibfnamefont {E.~N.}\ \bibnamefont {Hering}}, \bibinfo {author}
  {\bibfnamefont {S.~M.}\ \bibnamefont {Ramos}}, \bibinfo {author}
  {\bibfnamefont {M.~B.}\ \bibnamefont {Fontes}}, \bibinfo {author}
  {\bibfnamefont {E.}~\bibnamefont {Baggio-Saitovich}}, \bibinfo {author}
  {\bibfnamefont {H.}~\bibnamefont {Lee}}, \bibinfo {author} {\bibfnamefont
  {J.~L.}\ \bibnamefont {Sarrao}}, \bibinfo {author} {\bibfnamefont {J.~D.}\
  \bibnamefont {Thompson}},\ and\ \bibinfo {author} {\bibfnamefont {P.~G.}\
  \bibnamefont {Pagliuso}},\ }\bibfield  {title} {\bibinfo {title} {Tuning the
  pressure-induced superconducting phase in doped {{CeRhIn$_{5}$}}},\ }\href
  {https://doi.org/10.1103/PhysRevLett.101.017005} {\bibfield  {journal}
  {\bibinfo  {journal} {Phys. Rev. Lett.}\ }\textbf {\bibinfo {volume} {101}},\
  \bibinfo {pages} {017005} (\bibinfo {year} {2008})}\BibitemShut {NoStop}%
\bibitem [{\citenamefont {Knebel}\ \emph
  {et~al.}(2011{\natexlab{b}})\citenamefont {Knebel}, \citenamefont {Aoki},\
  and\ \citenamefont {Flouquet}}]{Knebel2011Comptes}%
  \BibitemOpen
  \bibfield  {author} {\bibinfo {author} {\bibfnamefont {G.}~\bibnamefont
  {Knebel}}, \bibinfo {author} {\bibfnamefont {D.}~\bibnamefont {Aoki}},\ and\
  \bibinfo {author} {\bibfnamefont {J.}~\bibnamefont {Flouquet}},\ }\bibfield
  {title} {\bibinfo {title} {Antiferromagnetism and superconductivity in cerium
  based heavy-fermion compounds},\ }\href
  {https://doi.org/10.1016/j.crhy.2011.05.002} {\bibfield  {journal} {\bibinfo
  {journal} {C. R. Phys.}\ }\textbf {\bibinfo {volume} {12}},\ \bibinfo {pages}
  {542} (\bibinfo {year} {2011}{\natexlab{b}})}\BibitemShut {NoStop}%
\bibitem [{\citenamefont {Seo}\ \emph {et~al.}(2015)\citenamefont {Seo},
  \citenamefont {Park}, \citenamefont {Bauer}, \citenamefont {Ronning},
  \citenamefont {Kim}, \citenamefont {Shim}, \citenamefont {Thompson},\ and\
  \citenamefont {Park}}]{Seo2015}%
  \BibitemOpen
  \bibfield  {author} {\bibinfo {author} {\bibfnamefont {S.}~\bibnamefont
  {Seo}}, \bibinfo {author} {\bibfnamefont {E.}~\bibnamefont {Park}}, \bibinfo
  {author} {\bibfnamefont {E.}~\bibnamefont {Bauer}}, \bibinfo {author}
  {\bibfnamefont {F.}~\bibnamefont {Ronning}}, \bibinfo {author} {\bibfnamefont
  {J.}~\bibnamefont {Kim}}, \bibinfo {author} {\bibfnamefont {J.-H.}\
  \bibnamefont {Shim}}, \bibinfo {author} {\bibfnamefont {J.}~\bibnamefont
  {Thompson}},\ and\ \bibinfo {author} {\bibfnamefont {T.}~\bibnamefont
  {Park}},\ }\bibfield  {title} {\bibinfo {title} {Controlling
  superconductivity by tunable quantum critical points},\ }\href
  {https://doi.org/doi.org/10.1038/ncomms7433} {\bibfield  {journal} {\bibinfo
  {journal} {Nat. Commun.}\ }\textbf {\bibinfo {volume} {6}},\ \bibinfo {pages}
  {1} (\bibinfo {year} {2015})}\BibitemShut {NoStop}%
\bibitem [{\citenamefont {Jung}\ \emph {et~al.}(2018)\citenamefont {Jung},
  \citenamefont {Seo}, \citenamefont {Lee}, \citenamefont {Bauer},
  \citenamefont {Lee},\ and\ \citenamefont {Park}}]{Jung2018}%
  \BibitemOpen
  \bibfield  {author} {\bibinfo {author} {\bibfnamefont {S.~G.}\ \bibnamefont
  {Jung}}, \bibinfo {author} {\bibfnamefont {S.}~\bibnamefont {Seo}}, \bibinfo
  {author} {\bibfnamefont {S.}~\bibnamefont {Lee}}, \bibinfo {author}
  {\bibfnamefont {E.~D.}\ \bibnamefont {Bauer}}, \bibinfo {author}
  {\bibfnamefont {H.-O.}\ \bibnamefont {Lee}},\ and\ \bibinfo {author}
  {\bibfnamefont {T.}~\bibnamefont {Park}},\ }\bibfield  {title} {\bibinfo
  {title} {A peak in the critical current for quantum critical
  superconductors},\ }\href {https://doi.org/10.1038/s41467-018-02899-5}
  {\bibfield  {journal} {\bibinfo  {journal} {Nat. Commun.}\ }\textbf {\bibinfo
  {volume} {9}},\ \bibinfo {pages} {1} (\bibinfo {year} {2018})}\BibitemShut
  {NoStop}%
\bibitem [{\citenamefont {Mathur}\ \emph {et~al.}(1998)\citenamefont {Mathur},
  \citenamefont {Grosche}, \citenamefont {Julian}, \citenamefont {Walker},
  \citenamefont {Freye}, \citenamefont {Haselwimmer},\ and\ \citenamefont
  {Lonzarich}}]{Mathur1998}%
  \BibitemOpen
  \bibfield  {author} {\bibinfo {author} {\bibfnamefont {N.}~\bibnamefont
  {Mathur}}, \bibinfo {author} {\bibfnamefont {F.}~\bibnamefont {Grosche}},
  \bibinfo {author} {\bibfnamefont {S.}~\bibnamefont {Julian}}, \bibinfo
  {author} {\bibfnamefont {I.}~\bibnamefont {Walker}}, \bibinfo {author}
  {\bibfnamefont {D.}~\bibnamefont {Freye}}, \bibinfo {author} {\bibfnamefont
  {R.}~\bibnamefont {Haselwimmer}},\ and\ \bibinfo {author} {\bibfnamefont
  {G.}~\bibnamefont {Lonzarich}},\ }\bibfield  {title} {\bibinfo {title}
  {Magnetically mediated superconductivity in heavy fermion compounds},\ }\href
  {https://doi.org/doi.org/10.1038/27838} {\bibfield  {journal} {\bibinfo
  {journal} {Nature}\ }\textbf {\bibinfo {volume} {394}},\ \bibinfo {pages}
  {39} (\bibinfo {year} {1998})}\BibitemShut {NoStop}%
\bibitem [{\citenamefont {Hegger}\ \emph {et~al.}(2000)\citenamefont {Hegger},
  \citenamefont {Petrovic}, \citenamefont {Moshopoulou}, \citenamefont
  {Hundley}, \citenamefont {Sarrao}, \citenamefont {Fisk},\ and\ \citenamefont
  {Thompson}}]{Hegger2000}%
  \BibitemOpen
  \bibfield  {author} {\bibinfo {author} {\bibfnamefont {H.}~\bibnamefont
  {Hegger}}, \bibinfo {author} {\bibfnamefont {C.}~\bibnamefont {Petrovic}},
  \bibinfo {author} {\bibfnamefont {E.~G.}\ \bibnamefont {Moshopoulou}},
  \bibinfo {author} {\bibfnamefont {M.~F.}\ \bibnamefont {Hundley}}, \bibinfo
  {author} {\bibfnamefont {J.~L.}\ \bibnamefont {Sarrao}}, \bibinfo {author}
  {\bibfnamefont {Z.}~\bibnamefont {Fisk}},\ and\ \bibinfo {author}
  {\bibfnamefont {J.~D.}\ \bibnamefont {Thompson}},\ }\bibfield  {title}
  {\bibinfo {title} {Pressure-induced superconductivity in quasi-2d
  {{CeRhIn$_5$}}},\ }\href {https://doi.org/10.1103/PhysRevLett.84.4986}
  {\bibfield  {journal} {\bibinfo  {journal} {Phys. Rev. Lett.}\ }\textbf
  {\bibinfo {volume} {84}},\ \bibinfo {pages} {4986} (\bibinfo {year}
  {2000})}\BibitemShut {NoStop}%
\bibitem [{\citenamefont {Shishido}\ \emph {et~al.}(2005)\citenamefont
  {Shishido}, \citenamefont {Settai}, \citenamefont {Harima},\ and\
  \citenamefont {{\=O}nuki}}]{Shishido2005}%
  \BibitemOpen
  \bibfield  {author} {\bibinfo {author} {\bibfnamefont {H.}~\bibnamefont
  {Shishido}}, \bibinfo {author} {\bibfnamefont {R.}~\bibnamefont {Settai}},
  \bibinfo {author} {\bibfnamefont {H.}~\bibnamefont {Harima}},\ and\ \bibinfo
  {author} {\bibfnamefont {Y.}~\bibnamefont {{\=O}nuki}},\ }\bibfield  {title}
  {\bibinfo {title} {A drastic change of the {{Fermi}} surface at a critical
  pressure in {{CeRhIn$_5$}}: {{dHvA}} study under pressure},\ }\href
  {https://doi.org/10.1143/JPSJ.74.1103} {\bibfield  {journal} {\bibinfo
  {journal} {J. Phys. Soc. Jpn.}\ }\textbf {\bibinfo {volume} {74}},\ \bibinfo
  {pages} {1103} (\bibinfo {year} {2005})}\BibitemShut {NoStop}%
\bibitem [{\citenamefont {Raymond}\ \emph {et~al.}(2014)\citenamefont
  {Raymond}, \citenamefont {Buhot}, \citenamefont {Ressouche}, \citenamefont
  {Bourdarot}, \citenamefont {Knebel},\ and\ \citenamefont
  {Lapertot}}]{Raymond2014}%
  \BibitemOpen
  \bibfield  {author} {\bibinfo {author} {\bibfnamefont {S.}~\bibnamefont
  {Raymond}}, \bibinfo {author} {\bibfnamefont {J.}~\bibnamefont {Buhot}},
  \bibinfo {author} {\bibfnamefont {E.}~\bibnamefont {Ressouche}}, \bibinfo
  {author} {\bibfnamefont {F.}~\bibnamefont {Bourdarot}}, \bibinfo {author}
  {\bibfnamefont {G.}~\bibnamefont {Knebel}},\ and\ \bibinfo {author}
  {\bibfnamefont {G.}~\bibnamefont {Lapertot}},\ }\bibfield  {title} {\bibinfo
  {title} {Switching of the magnetic order in {{CeRhIn$_{5-x}$Sn$_{x}$}} in the
  vicinity of its quantum critical point},\ }\href
  {https://doi.org/10.1103/PhysRevB.90.014423} {\bibfield  {journal} {\bibinfo
  {journal} {Phys. Rev. B}\ }\textbf {\bibinfo {volume} {90}},\ \bibinfo
  {pages} {014423} (\bibinfo {year} {2014})}\BibitemShut {NoStop}%
\bibitem [{\citenamefont {Bauer}\ \emph {et~al.}(2006)\citenamefont {Bauer},
  \citenamefont {Mixson}, \citenamefont {Ronning}, \citenamefont {Hur},
  \citenamefont {Movshovich}, \citenamefont {Thompson}, \citenamefont {Sarrao},
  \citenamefont {Hundley}, \citenamefont {Tobash},\ and\ \citenamefont
  {Bobev}}]{Bauer2006}%
  \BibitemOpen
  \bibfield  {author} {\bibinfo {author} {\bibfnamefont {E.}~\bibnamefont
  {Bauer}}, \bibinfo {author} {\bibfnamefont {D.}~\bibnamefont {Mixson}},
  \bibinfo {author} {\bibfnamefont {F.}~\bibnamefont {Ronning}}, \bibinfo
  {author} {\bibfnamefont {N.}~\bibnamefont {Hur}}, \bibinfo {author}
  {\bibfnamefont {R.}~\bibnamefont {Movshovich}}, \bibinfo {author}
  {\bibfnamefont {J.}~\bibnamefont {Thompson}}, \bibinfo {author}
  {\bibfnamefont {J.}~\bibnamefont {Sarrao}}, \bibinfo {author} {\bibfnamefont
  {M.}~\bibnamefont {Hundley}}, \bibinfo {author} {\bibfnamefont
  {P.}~\bibnamefont {Tobash}},\ and\ \bibinfo {author} {\bibfnamefont
  {S.}~\bibnamefont {Bobev}},\ }\bibfield  {title} {\bibinfo {title}
  {Antiferromagnetic quantum critical point in {{CeRhIn$_{5-x}$Sn$_x$}}},\
  }\href {https://doi.org/10.1016/j.physb.2006.01.053} {\bibfield  {journal}
  {\bibinfo  {journal} {Physica B}\ }\textbf {\bibinfo {volume} {378-380}},\
  \bibinfo {pages} {142} (\bibinfo {year} {2006})}\BibitemShut {NoStop}%
\bibitem [{\citenamefont {Park}\ \emph {et~al.}(2020)\citenamefont {Park},
  \citenamefont {Shin}, \citenamefont {Lee}, \citenamefont {Seo}, \citenamefont
  {Jang}, \citenamefont {Kim}, \citenamefont {Lee}, \citenamefont {Wang},
  \citenamefont {Lee},\ and\ \citenamefont {Park}}]{Park2020}%
  \BibitemOpen
  \bibfield  {author} {\bibinfo {author} {\bibfnamefont {T.~B.}\ \bibnamefont
  {Park}}, \bibinfo {author} {\bibfnamefont {S.}~\bibnamefont {Shin}}, \bibinfo
  {author} {\bibfnamefont {S.}~\bibnamefont {Lee}}, \bibinfo {author}
  {\bibfnamefont {S.}~\bibnamefont {Seo}}, \bibinfo {author} {\bibfnamefont
  {H.}~\bibnamefont {Jang}}, \bibinfo {author} {\bibfnamefont {J.}~\bibnamefont
  {Kim}}, \bibinfo {author} {\bibfnamefont {H.}~\bibnamefont {Lee}}, \bibinfo
  {author} {\bibfnamefont {H.}~\bibnamefont {Wang}}, \bibinfo {author}
  {\bibfnamefont {H.}~\bibnamefont {Lee}},\ and\ \bibinfo {author}
  {\bibfnamefont {T.}~\bibnamefont {Park}},\ }\bibfield  {title} {\bibinfo
  {title} {Evolution of antiferromagnetism in {{Zn}}-doped heavy-fermion
  compound {{CeRh(In$_{1-x}$Zn$_x$)$_5$}}},\ }\href
  {https://doi.org/10.1103/PhysRevMaterials.4.084801} {\bibfield  {journal}
  {\bibinfo  {journal} {Phys. Rev. Mater.}\ }\textbf {\bibinfo {volume} {4}},\
  \bibinfo {pages} {084801} (\bibinfo {year} {2020})}\BibitemShut {NoStop}%
\bibitem [{\citenamefont {Donath}\ \emph {et~al.}(2009)\citenamefont {Donath},
  \citenamefont {Steglich}, \citenamefont {Bauer}, \citenamefont {Ronning},
  \citenamefont {Sarrao},\ and\ \citenamefont {Gegenwart}}]{Donath2009}%
  \BibitemOpen
  \bibfield  {author} {\bibinfo {author} {\bibfnamefont {J.~G.}\ \bibnamefont
  {Donath}}, \bibinfo {author} {\bibfnamefont {F.}~\bibnamefont {Steglich}},
  \bibinfo {author} {\bibfnamefont {E.~D.}\ \bibnamefont {Bauer}}, \bibinfo
  {author} {\bibfnamefont {F.}~\bibnamefont {Ronning}}, \bibinfo {author}
  {\bibfnamefont {J.~L.}\ \bibnamefont {Sarrao}},\ and\ \bibinfo {author}
  {\bibfnamefont {P.}~\bibnamefont {Gegenwart}},\ }\bibfield  {title} {\bibinfo
  {title} {Quantum criticality in layered {{CeRhIn$_{5-x}$Sn$_x$}} compared
  with cubic {{CeIn$_{3-x}$Sn$_x$}}},\ }\href
  {https://doi.org/10.1209/0295-5075/87/57011} {\bibfield  {journal} {\bibinfo
  {journal} {Europhys. Lett.}\ }\textbf {\bibinfo {volume} {87}},\ \bibinfo
  {pages} {57011} (\bibinfo {year} {2009})}\BibitemShut {NoStop}%
\bibitem [{\citenamefont {Basov}\ \emph {et~al.}(2011)\citenamefont {Basov},
  \citenamefont {Averitt}, \citenamefont {van~der Marel}, \citenamefont
  {Dressel},\ and\ \citenamefont {Haule}}]{Basov2011}%
  \BibitemOpen
  \bibfield  {author} {\bibinfo {author} {\bibfnamefont {D.~N.}\ \bibnamefont
  {Basov}}, \bibinfo {author} {\bibfnamefont {R.~D.}\ \bibnamefont {Averitt}},
  \bibinfo {author} {\bibfnamefont {D.}~\bibnamefont {van~der Marel}}, \bibinfo
  {author} {\bibfnamefont {M.}~\bibnamefont {Dressel}},\ and\ \bibinfo {author}
  {\bibfnamefont {K.}~\bibnamefont {Haule}},\ }\bibfield  {title} {\bibinfo
  {title} {Electrodynamics of correlated electron materials},\ }\href
  {https://doi.org/10.1103/RevModPhys.83.471} {\bibfield  {journal} {\bibinfo
  {journal} {Rev. Mod. Phys.}\ }\textbf {\bibinfo {volume} {83}},\ \bibinfo
  {pages} {471} (\bibinfo {year} {2011})}\BibitemShut {NoStop}%
\bibitem [{\citenamefont {Kimura}\ and\ \citenamefont
  {Okamura}(2012)}]{Kimura2012}%
  \BibitemOpen
  \bibfield  {author} {\bibinfo {author} {\bibfnamefont {S.}~\bibnamefont
  {Kimura}}\ and\ \bibinfo {author} {\bibfnamefont {H.}~\bibnamefont
  {Okamura}},\ }\bibfield  {title} {\bibinfo {title} {Infrared and terahertz
  spectroscopy of strongly correlated electron systems under extreme
  conditions},\ }\href {https://doi.org/10.7566/JPSJ.82.021004} {\bibfield
  {journal} {\bibinfo  {journal} {J. Phys. Soc. Jpn.}\ }\textbf {\bibinfo
  {volume} {82}},\ \bibinfo {pages} {021004} (\bibinfo {year}
  {2012})}\BibitemShut {NoStop}%
\bibitem [{\citenamefont {Chen}\ and\ \citenamefont {Wang}(2016)}]{Chen2016}%
  \BibitemOpen
  \bibfield  {author} {\bibinfo {author} {\bibfnamefont {R.~Y.}\ \bibnamefont
  {Chen}}\ and\ \bibinfo {author} {\bibfnamefont {N.~L.}\ \bibnamefont
  {Wang}},\ }\bibfield  {title} {\bibinfo {title} {Infrared properties of heavy
  fermions: evolution from weak to strong hybridizations},\ }\href
  {https://doi.org/10.1088/0034-4885/79/6/064502} {\bibfield  {journal}
  {\bibinfo  {journal} {Rep. Progr. Phys.}\ }\textbf {\bibinfo {volume} {79}},\
  \bibinfo {pages} {064502} (\bibinfo {year} {2016})}\BibitemShut {NoStop}%
\bibitem [{\citenamefont {Dressel}\ and\ \citenamefont
  {Gr{\"u}ner}(2002)}]{DresselGruner2002}%
  \BibitemOpen
  \bibfield  {author} {\bibinfo {author} {\bibfnamefont {M.}~\bibnamefont
  {Dressel}}\ and\ \bibinfo {author} {\bibfnamefont {G.}~\bibnamefont
  {Gr{\"u}ner}},\ }\href@noop {} {\emph {\bibinfo {title} {Electrodynamics of
  Solids}}}\ (\bibinfo  {publisher} {Cambridge University Press},\ \bibinfo
  {address} {Cambridge},\ \bibinfo {year} {2002})\BibitemShut {NoStop}%
\bibitem [{\citenamefont {Degiorgi}(1999)}]{Degiorgi1999}%
  \BibitemOpen
  \bibfield  {author} {\bibinfo {author} {\bibfnamefont {L.}~\bibnamefont
  {Degiorgi}},\ }\bibfield  {title} {\bibinfo {title} {The electrodynamic
  response of heavy-electron compounds},\ }\href
  {https://doi.org/10.1103/RevModPhys.71.687} {\bibfield  {journal} {\bibinfo
  {journal} {Rev. Mod. Phys.}\ }\textbf {\bibinfo {volume} {71}},\ \bibinfo
  {pages} {687} (\bibinfo {year} {1999})}\BibitemShut {NoStop}%
\bibitem [{\citenamefont {Kimura}\ \emph {et~al.}(2016)\citenamefont {Kimura},
  \citenamefont {Kwon}, \citenamefont {Matsumoto}, \citenamefont {Aoki},\ and\
  \citenamefont {Sakai}}]{Kimura2016}%
  \BibitemOpen
  \bibfield  {author} {\bibinfo {author} {\bibfnamefont {S.}~\bibnamefont
  {Kimura}}, \bibinfo {author} {\bibfnamefont {Y.~S.}\ \bibnamefont {Kwon}},
  \bibinfo {author} {\bibfnamefont {Y.}~\bibnamefont {Matsumoto}}, \bibinfo
  {author} {\bibfnamefont {H.}~\bibnamefont {Aoki}},\ and\ \bibinfo {author}
  {\bibfnamefont {O.}~\bibnamefont {Sakai}},\ }\bibfield  {title} {\bibinfo
  {title} {Optical evidence of itinerant-localized crossover of 4$f$ electrons
  in cerium compounds},\ }\href {https://doi.org/10.7566/JPSJ.85.083702}
  {\bibfield  {journal} {\bibinfo  {journal} {J. Phys. Soc. Jpn.}\ }\textbf
  {\bibinfo {volume} {85}},\ \bibinfo {pages} {083702} (\bibinfo {year}
  {2016})}\BibitemShut {NoStop}%
\bibitem [{\citenamefont {Okamura}\ \emph {et~al.}(2007)\citenamefont
  {Okamura}, \citenamefont {Watanabe}, \citenamefont {Matsunami}, \citenamefont
  {Nishihara}, \citenamefont {Tsujii}, \citenamefont {Ebihara}, \citenamefont
  {Sugawara}, \citenamefont {Sato}, \citenamefont {{\=O}nuki}, \citenamefont
  {Isikawa} \emph {et~al.}}]{Okamura2007}%
  \BibitemOpen
  \bibfield  {author} {\bibinfo {author} {\bibfnamefont {H.}~\bibnamefont
  {Okamura}}, \bibinfo {author} {\bibfnamefont {T.}~\bibnamefont {Watanabe}},
  \bibinfo {author} {\bibfnamefont {M.}~\bibnamefont {Matsunami}}, \bibinfo
  {author} {\bibfnamefont {T.}~\bibnamefont {Nishihara}}, \bibinfo {author}
  {\bibfnamefont {N.}~\bibnamefont {Tsujii}}, \bibinfo {author} {\bibfnamefont
  {T.}~\bibnamefont {Ebihara}}, \bibinfo {author} {\bibfnamefont
  {H.}~\bibnamefont {Sugawara}}, \bibinfo {author} {\bibfnamefont
  {H.}~\bibnamefont {Sato}}, \bibinfo {author} {\bibfnamefont {Y.}~\bibnamefont
  {{\=O}nuki}}, \bibinfo {author} {\bibfnamefont {Y.}~\bibnamefont {Isikawa}},
  \emph {et~al.},\ }\bibfield  {title} {\bibinfo {title} {Universal scaling in
  the dynamical conductivity of heavy fermion {{Ce}} and {{Yb}} compounds},\
  }\href {https://doi.org/10.1143/JPSJ.76.023703} {\bibfield  {journal}
  {\bibinfo  {journal} {J. Phys. Soc. Jpn.}\ }\textbf {\bibinfo {volume}
  {76}},\ \bibinfo {pages} {023703} (\bibinfo {year} {2007})}\BibitemShut
  {NoStop}%
\bibitem [{\citenamefont {Dordevic}\ \emph {et~al.}(2001)\citenamefont
  {Dordevic}, \citenamefont {Basov}, \citenamefont {Dilley}, \citenamefont
  {Bauer},\ and\ \citenamefont {Maple}}]{Dordevic2001}%
  \BibitemOpen
  \bibfield  {author} {\bibinfo {author} {\bibfnamefont {S.~V.}\ \bibnamefont
  {Dordevic}}, \bibinfo {author} {\bibfnamefont {D.~N.}\ \bibnamefont {Basov}},
  \bibinfo {author} {\bibfnamefont {N.~R.}\ \bibnamefont {Dilley}}, \bibinfo
  {author} {\bibfnamefont {E.~D.}\ \bibnamefont {Bauer}},\ and\ \bibinfo
  {author} {\bibfnamefont {M.~B.}\ \bibnamefont {Maple}},\ }\bibfield  {title}
  {\bibinfo {title} {Hybridization gap in heavy fermion compounds},\ }\href
  {https://doi.org/10.1103/PhysRevLett.86.684} {\bibfield  {journal} {\bibinfo
  {journal} {Phys. Rev. Lett.}\ }\textbf {\bibinfo {volume} {86}},\ \bibinfo
  {pages} {684} (\bibinfo {year} {2001})}\BibitemShut {NoStop}%
\bibitem [{\citenamefont {Scheffler}\ \emph {et~al.}(2005)\citenamefont
  {Scheffler}, \citenamefont {Dressel}, \citenamefont {Jourdan},\ and\
  \citenamefont {Adrian}}]{Scheffler2005}%
  \BibitemOpen
  \bibfield  {author} {\bibinfo {author} {\bibfnamefont {M.}~\bibnamefont
  {Scheffler}}, \bibinfo {author} {\bibfnamefont {M.}~\bibnamefont {Dressel}},
  \bibinfo {author} {\bibfnamefont {M.}~\bibnamefont {Jourdan}},\ and\ \bibinfo
  {author} {\bibfnamefont {H.}~\bibnamefont {Adrian}},\ }\bibfield  {title}
  {\bibinfo {title} {Extremely slow {{Drude}} relaxation of correlated
  electrons},\ }\href {https://doi.org/10.1038/nature04232} {\bibfield
  {journal} {\bibinfo  {journal} {Nature}\ }\textbf {\bibinfo {volume} {438}},\
  \bibinfo {pages} {1135} (\bibinfo {year} {2005})}\BibitemShut {NoStop}%
\bibitem [{\citenamefont {Kimura}\ \emph {et~al.}(2006)\citenamefont {Kimura},
  \citenamefont {Sichelschmidt}, \citenamefont {Ferstl}, \citenamefont
  {Krellner}, \citenamefont {Geibel},\ and\ \citenamefont
  {Steglich}}]{Kimura2006}%
  \BibitemOpen
  \bibfield  {author} {\bibinfo {author} {\bibfnamefont {S.}~\bibnamefont
  {Kimura}}, \bibinfo {author} {\bibfnamefont {J.}~\bibnamefont
  {Sichelschmidt}}, \bibinfo {author} {\bibfnamefont {J.}~\bibnamefont
  {Ferstl}}, \bibinfo {author} {\bibfnamefont {C.}~\bibnamefont {Krellner}},
  \bibinfo {author} {\bibfnamefont {C.}~\bibnamefont {Geibel}},\ and\ \bibinfo
  {author} {\bibfnamefont {F.}~\bibnamefont {Steglich}},\ }\bibfield  {title}
  {\bibinfo {title} {Optical observation of non-{{Fermi}}-liquid behavior in
  the heavy fermion state of {{YbRh$_2$Si$_2$}}},\ }\href
  {https://doi.org/10.1103/PhysRevB.74.132408} {\bibfield  {journal} {\bibinfo
  {journal} {Phys. Rev. B.}\ }\textbf {\bibinfo {volume} {74}},\ \bibinfo
  {pages} {132408} (\bibinfo {year} {2006})}\BibitemShut {NoStop}%
\bibitem [{\citenamefont {Scheffler}\ \emph {et~al.}(2013)\citenamefont
  {Scheffler}, \citenamefont {Schlegel}, \citenamefont {Clauss}, \citenamefont
  {Hafner}, \citenamefont {Fella}, \citenamefont {Dressel}, \citenamefont
  {Jourdan}, \citenamefont {Sichelschmidt}, \citenamefont {Krellner},
  \citenamefont {Geibel},\ and\ \citenamefont {Steglich}}]{Scheffler2013}%
  \BibitemOpen
  \bibfield  {author} {\bibinfo {author} {\bibfnamefont {M.}~\bibnamefont
  {Scheffler}}, \bibinfo {author} {\bibfnamefont {K.}~\bibnamefont {Schlegel}},
  \bibinfo {author} {\bibfnamefont {C.}~\bibnamefont {Clauss}}, \bibinfo
  {author} {\bibfnamefont {D.}~\bibnamefont {Hafner}}, \bibinfo {author}
  {\bibfnamefont {C.}~\bibnamefont {Fella}}, \bibinfo {author} {\bibfnamefont
  {M.}~\bibnamefont {Dressel}}, \bibinfo {author} {\bibfnamefont
  {M.}~\bibnamefont {Jourdan}}, \bibinfo {author} {\bibfnamefont
  {J.}~\bibnamefont {Sichelschmidt}}, \bibinfo {author} {\bibfnamefont
  {C.}~\bibnamefont {Krellner}}, \bibinfo {author} {\bibfnamefont
  {C.}~\bibnamefont {Geibel}},\ and\ \bibinfo {author} {\bibfnamefont
  {F.}~\bibnamefont {Steglich}},\ }\bibfield  {title} {\bibinfo {title}
  {Microwave spectroscopy on heavy-fermion systems: Probing the dynamics of
  charges and magnetic moments},\ }\href
  {https://doi.org/10.1002/pssb.201200925} {\bibfield  {journal} {\bibinfo
  {journal} {phys. stat. sol. (b)}\ }\textbf {\bibinfo {volume} {250}},\
  \bibinfo {pages} {439} (\bibinfo {year} {2013})}\BibitemShut {NoStop}%
\bibitem [{\citenamefont {Dressel}\ \emph
  {et~al.}(2002{\natexlab{a}})\citenamefont {Dressel}, \citenamefont {Kasper},
  \citenamefont {Petukhov}, \citenamefont {Gorshunov}, \citenamefont
  {Gr{\"u}ner}, \citenamefont {Huth},\ and\ \citenamefont
  {Adrian}}]{Dressel2002Letter}%
  \BibitemOpen
  \bibfield  {author} {\bibinfo {author} {\bibfnamefont {M.}~\bibnamefont
  {Dressel}}, \bibinfo {author} {\bibfnamefont {N.}~\bibnamefont {Kasper}},
  \bibinfo {author} {\bibfnamefont {K.}~\bibnamefont {Petukhov}}, \bibinfo
  {author} {\bibfnamefont {B.}~\bibnamefont {Gorshunov}}, \bibinfo {author}
  {\bibfnamefont {G.}~\bibnamefont {Gr{\"u}ner}}, \bibinfo {author}
  {\bibfnamefont {M.}~\bibnamefont {Huth}},\ and\ \bibinfo {author}
  {\bibfnamefont {H.}~\bibnamefont {Adrian}},\ }\bibfield  {title} {\bibinfo
  {title} {Nature of heavy quasiparticles in magnetically ordered heavy
  fermions {{UPd$_2$Al$_3$}} and {{UPt$_3$}}},\ }\href
  {https://doi.org/10.1103/PhysRevLett.88.186404} {\bibfield  {journal}
  {\bibinfo  {journal} {Phys. Rev. Lett.}\ }\textbf {\bibinfo {volume} {88}},\
  \bibinfo {pages} {186404} (\bibinfo {year} {2002}{\natexlab{a}})}\BibitemShut
  {NoStop}%
\bibitem [{\citenamefont {Dressel}\ \emph
  {et~al.}(2002{\natexlab{b}})\citenamefont {Dressel}, \citenamefont {Kasper},
  \citenamefont {Petukhov}, \citenamefont {Peligrad}, \citenamefont
  {Gorshunov}, \citenamefont {Jourdan}, \citenamefont {Huth},\ and\
  \citenamefont {Adrian}}]{Dressel2002}%
  \BibitemOpen
  \bibfield  {author} {\bibinfo {author} {\bibfnamefont {M.}~\bibnamefont
  {Dressel}}, \bibinfo {author} {\bibfnamefont {N.}~\bibnamefont {Kasper}},
  \bibinfo {author} {\bibfnamefont {K.}~\bibnamefont {Petukhov}}, \bibinfo
  {author} {\bibfnamefont {D.~N.}\ \bibnamefont {Peligrad}}, \bibinfo {author}
  {\bibfnamefont {B.}~\bibnamefont {Gorshunov}}, \bibinfo {author}
  {\bibfnamefont {M.}~\bibnamefont {Jourdan}}, \bibinfo {author} {\bibfnamefont
  {M.}~\bibnamefont {Huth}},\ and\ \bibinfo {author} {\bibfnamefont
  {H.}~\bibnamefont {Adrian}},\ }\bibfield  {title} {\bibinfo {title}
  {Correlation gap in the heavy-fermion antiferromagnet {{UPd$_2$Al$_3$}}},\
  }\href {https://doi.org/10.1103/PhysRevB.66.035110} {\bibfield  {journal}
  {\bibinfo  {journal} {Phys. Rev. B.}\ }\textbf {\bibinfo {volume} {66}},\
  \bibinfo {pages} {035110} (\bibinfo {year} {2002}{\natexlab{b}})}\BibitemShut
  {NoStop}%
\bibitem [{\citenamefont {Lee}\ \emph {et~al.}(2008)\citenamefont {Lee},
  \citenamefont {Lee}, \citenamefont {Oh}, \citenamefont {Im}, \citenamefont
  {Park}, \citenamefont {Kimura},\ and\ \citenamefont {Kwon}}]{Lee2008}%
  \BibitemOpen
  \bibfield  {author} {\bibinfo {author} {\bibfnamefont {K.}~\bibnamefont
  {Lee}}, \bibinfo {author} {\bibfnamefont {C.}~\bibnamefont {Lee}}, \bibinfo
  {author} {\bibfnamefont {H.}~\bibnamefont {Oh}}, \bibinfo {author}
  {\bibfnamefont {H.}~\bibnamefont {Im}}, \bibinfo {author} {\bibfnamefont
  {T.}~\bibnamefont {Park}}, \bibinfo {author} {\bibfnamefont {S.}~\bibnamefont
  {Kimura}},\ and\ \bibinfo {author} {\bibfnamefont {Y.}~\bibnamefont {Kwon}},\
  }\bibfield  {title} {\bibinfo {title} {Optical evidence for a change in the
  heavy electron {{Fermi}} surface at a magnetic quantum critical point of
  {{CeNi$_{1-x}$Co$_{x}$Ge$_{2}$}}},\ }\href
  {https://doi.org/10.1088/0953-8984/20/28/285202} {\bibfield  {journal}
  {\bibinfo  {journal} {J. Phys.: Condens. Matter}\ }\textbf {\bibinfo {volume}
  {20}},\ \bibinfo {pages} {285202} (\bibinfo {year} {2008})}\BibitemShut
  {NoStop}%
\bibitem [{\citenamefont {Tanner}(2015)}]{Tanner2015}%
  \BibitemOpen
  \bibfield  {author} {\bibinfo {author} {\bibfnamefont {D.~B.}\ \bibnamefont
  {Tanner}},\ }\bibfield  {title} {\bibinfo {title} {Use of x-ray scattering
  functions in kramers-kronig analysis of reflectance},\ }\href
  {https://doi.org/10.1103/PhysRevB.91.035123} {\bibfield  {journal} {\bibinfo
  {journal} {Phys. Rev. B.}\ }\textbf {\bibinfo {volume} {91}},\ \bibinfo
  {pages} {035123} (\bibinfo {year} {2015})}\BibitemShut {NoStop}%
\bibitem [{\citenamefont {Mena}\ \emph {et~al.}(2005)\citenamefont {Mena},
  \citenamefont {van~der Marel},\ and\ \citenamefont {Sarrao}}]{Mena2005}%
  \BibitemOpen
  \bibfield  {author} {\bibinfo {author} {\bibfnamefont {F.~P.}\ \bibnamefont
  {Mena}}, \bibinfo {author} {\bibfnamefont {D.}~\bibnamefont {van~der
  Marel}},\ and\ \bibinfo {author} {\bibfnamefont {J.~L.}\ \bibnamefont
  {Sarrao}},\ }\bibfield  {title} {\bibinfo {title} {Optical conductivity of
  {{Ce$M$In$ _5$}} ({{$M$}} = {{Co}}, {{Rh}}, {{Ir}})},\ }\href
  {https://doi.org/10.1103/PhysRevB.72.045119} {\bibfield  {journal} {\bibinfo
  {journal} {Phys. Rev. B.}\ }\textbf {\bibinfo {volume} {72}},\ \bibinfo
  {pages} {045119} (\bibinfo {year} {2005})}\BibitemShut {NoStop}%
\bibitem [{\citenamefont {Burch}\ \emph {et~al.}(2007)\citenamefont {Burch},
  \citenamefont {Dordevic}, \citenamefont {Mena}, \citenamefont {Kuzmenko},
  \citenamefont {van~der Marel}, \citenamefont {Sarrao}, \citenamefont
  {Jeffries}, \citenamefont {Bauer}, \citenamefont {Maple},\ and\ \citenamefont
  {Basov}}]{Burch2007}%
  \BibitemOpen
  \bibfield  {author} {\bibinfo {author} {\bibfnamefont {K.~S.}\ \bibnamefont
  {Burch}}, \bibinfo {author} {\bibfnamefont {S.~V.}\ \bibnamefont {Dordevic}},
  \bibinfo {author} {\bibfnamefont {F.~P.}\ \bibnamefont {Mena}}, \bibinfo
  {author} {\bibfnamefont {A.~B.}\ \bibnamefont {Kuzmenko}}, \bibinfo {author}
  {\bibfnamefont {D.}~\bibnamefont {van~der Marel}}, \bibinfo {author}
  {\bibfnamefont {J.~L.}\ \bibnamefont {Sarrao}}, \bibinfo {author}
  {\bibfnamefont {J.~R.}\ \bibnamefont {Jeffries}}, \bibinfo {author}
  {\bibfnamefont {E.~D.}\ \bibnamefont {Bauer}}, \bibinfo {author}
  {\bibfnamefont {M.~B.}\ \bibnamefont {Maple}},\ and\ \bibinfo {author}
  {\bibfnamefont {D.~N.}\ \bibnamefont {Basov}},\ }\bibfield  {title} {\bibinfo
  {title} {Optical signatures of momentum-dependent hybridization of the local
  moments and conduction electrons in {{Kondo}} lattices},\ }\href
  {https://doi.org/10.1103/PhysRevB.75.054523} {\bibfield  {journal} {\bibinfo
  {journal} {Phys. Rev. B.}\ }\textbf {\bibinfo {volume} {75}},\ \bibinfo
  {pages} {054523} (\bibinfo {year} {2007})}\BibitemShut {NoStop}%
\bibitem [{\citenamefont {Okamura}\ \emph {et~al.}(2015)\citenamefont
  {Okamura}, \citenamefont {Takigawa}, \citenamefont {Bauer}, \citenamefont
  {Moriwaki},\ and\ \citenamefont {Ikemoto}}]{Okamura2015}%
  \BibitemOpen
  \bibfield  {author} {\bibinfo {author} {\bibfnamefont {H.}~\bibnamefont
  {Okamura}}, \bibinfo {author} {\bibfnamefont {A.}~\bibnamefont {Takigawa}},
  \bibinfo {author} {\bibfnamefont {E.}~\bibnamefont {Bauer}}, \bibinfo
  {author} {\bibfnamefont {T.}~\bibnamefont {Moriwaki}},\ and\ \bibinfo
  {author} {\bibfnamefont {Y.}~\bibnamefont {Ikemoto}},\ }\bibfield  {title}
  {\bibinfo {title} {Pressure evolution of $f$ electron hybridized state in
  {{CeCoIn$_5$}} studied by optical conductivity},\ }\href
  {https://doi.org/10.1088/1742-6596/592/1/012001} {\bibfield  {journal}
  {\bibinfo  {journal} {J. Phys. Conf. Ser.}\ }\textbf {\bibinfo {volume}
  {592}},\ \bibinfo {pages} {012001} (\bibinfo {year} {2015})}\BibitemShut
  {NoStop}%
\bibitem [{\citenamefont {Iizuka}\ \emph {et~al.}(2012)\citenamefont {Iizuka},
  \citenamefont {Mizuno}, \citenamefont {Hun~Min}, \citenamefont {Seung~Kwon},\
  and\ \citenamefont {Kimura}}]{Iizuka2012}%
  \BibitemOpen
  \bibfield  {author} {\bibinfo {author} {\bibfnamefont {T.}~\bibnamefont
  {Iizuka}}, \bibinfo {author} {\bibfnamefont {T.}~\bibnamefont {Mizuno}},
  \bibinfo {author} {\bibfnamefont {B.}~\bibnamefont {Hun~Min}}, \bibinfo
  {author} {\bibfnamefont {Y.}~\bibnamefont {Seung~Kwon}},\ and\ \bibinfo
  {author} {\bibfnamefont {S.-i.}\ \bibnamefont {Kimura}},\ }\bibfield  {title}
  {\bibinfo {title} {Existence of heavy fermions in the antiferromagnetic phase
  of {{CeIn$_3$}}},\ }\href {https://doi.org/10.1143/JPSJ.81.043703} {\bibfield
   {journal} {\bibinfo  {journal} {J. Phys. Soc. Jpn.}\ }\textbf {\bibinfo
  {volume} {81}},\ \bibinfo {pages} {043703} (\bibinfo {year}
  {2012})}\BibitemShut {NoStop}%
\bibitem [{\citenamefont {Singley}\ \emph {et~al.}(2002)\citenamefont
  {Singley}, \citenamefont {Basov}, \citenamefont {Bauer},\ and\ \citenamefont
  {Maple}}]{Singley2002}%
  \BibitemOpen
  \bibfield  {author} {\bibinfo {author} {\bibfnamefont {E.~J.}\ \bibnamefont
  {Singley}}, \bibinfo {author} {\bibfnamefont {D.~N.}\ \bibnamefont {Basov}},
  \bibinfo {author} {\bibfnamefont {E.~D.}\ \bibnamefont {Bauer}},\ and\
  \bibinfo {author} {\bibfnamefont {M.~B.}\ \bibnamefont {Maple}},\ }\bibfield
  {title} {\bibinfo {title} {Optical conductivity of the heavy fermion
  superconductor {{CeCoIn$_5$}}},\ }\href
  {https://doi.org/10.1103/PhysRevB.65.161101} {\bibfield  {journal} {\bibinfo
  {journal} {Phys. Rev. B.}\ }\textbf {\bibinfo {volume} {65}},\ \bibinfo
  {pages} {161101} (\bibinfo {year} {2002})}\BibitemShut {NoStop}%
\bibitem [{\citenamefont {Okamura}\ \emph {et~al.}(2019)\citenamefont
  {Okamura}, \citenamefont {Takigawa}, \citenamefont {Yamasaki}, \citenamefont
  {Bauer}, \citenamefont {Ohara}, \citenamefont {Ikemoto},\ and\ \citenamefont
  {Moriwaki}}]{Okamura2019}%
  \BibitemOpen
  \bibfield  {author} {\bibinfo {author} {\bibfnamefont {H.}~\bibnamefont
  {Okamura}}, \bibinfo {author} {\bibfnamefont {A.}~\bibnamefont {Takigawa}},
  \bibinfo {author} {\bibfnamefont {T.}~\bibnamefont {Yamasaki}}, \bibinfo
  {author} {\bibfnamefont {E.~D.}\ \bibnamefont {Bauer}}, \bibinfo {author}
  {\bibfnamefont {S.}~\bibnamefont {Ohara}}, \bibinfo {author} {\bibfnamefont
  {Y.}~\bibnamefont {Ikemoto}},\ and\ \bibinfo {author} {\bibfnamefont
  {T.}~\bibnamefont {Moriwaki}},\ }\bibfield  {title} {\bibinfo {title}
  {Contrasting pressure evolution of $f$-electron hybridized states in
  {{CeRhIn$_5$}} and {{YbNi$_3$Ga$_9$}}: An optical conductivity study},\
  }\href {https://doi.org/10.1103/PhysRevB.100.195112} {\bibfield  {journal}
  {\bibinfo  {journal} {Phys. Rev. B.}\ }\textbf {\bibinfo {volume} {100}},\
  \bibinfo {pages} {195112(R)} (\bibinfo {year} {2019})}\BibitemShut {NoStop}%
\bibitem [{\citenamefont {Luo}\ \emph {et~al.}(2014)\citenamefont {Luo},
  \citenamefont {Pourovskii}, \citenamefont {Rowley}, \citenamefont {Li},
  \citenamefont {Feng}, \citenamefont {Georges}, \citenamefont {Dai},
  \citenamefont {Cao}, \citenamefont {Xu}, \citenamefont {Si} \emph
  {et~al.}}]{Luo2014}%
  \BibitemOpen
  \bibfield  {author} {\bibinfo {author} {\bibfnamefont {Y.}~\bibnamefont
  {Luo}}, \bibinfo {author} {\bibfnamefont {L.}~\bibnamefont {Pourovskii}},
  \bibinfo {author} {\bibfnamefont {S.}~\bibnamefont {Rowley}}, \bibinfo
  {author} {\bibfnamefont {Y.}~\bibnamefont {Li}}, \bibinfo {author}
  {\bibfnamefont {C.}~\bibnamefont {Feng}}, \bibinfo {author} {\bibfnamefont
  {A.}~\bibnamefont {Georges}}, \bibinfo {author} {\bibfnamefont
  {J.}~\bibnamefont {Dai}}, \bibinfo {author} {\bibfnamefont {G.}~\bibnamefont
  {Cao}}, \bibinfo {author} {\bibfnamefont {Z.}~\bibnamefont {Xu}}, \bibinfo
  {author} {\bibfnamefont {Q.}~\bibnamefont {Si}}, \emph {et~al.},\ }\bibfield
  {title} {\bibinfo {title} {Heavy-fermion quantum criticality and destruction
  of the {{Kondo}} effect in a nickel oxypnictide},\ }\href
  {https://doi.org/doi.org/10.1038/nmat3991} {\bibfield  {journal} {\bibinfo
  {journal} {Nat. Mater.}\ }\textbf {\bibinfo {volume} {13}},\ \bibinfo {pages}
  {777} (\bibinfo {year} {2014})}\BibitemShut {NoStop}%
\bibitem [{\citenamefont {Friedemann}\ \emph {et~al.}(2009)\citenamefont
  {Friedemann}, \citenamefont {Westerkamp}, \citenamefont {Brando},
  \citenamefont {Oeschler}, \citenamefont {Wirth}, \citenamefont {Krellner},
  \citenamefont {Geibel},\ and\ \citenamefont {Steglich}}]{Friedemann2009}%
  \BibitemOpen
  \bibfield  {author} {\bibinfo {author} {\bibfnamefont {S.}~\bibnamefont
  {Friedemann}}, \bibinfo {author} {\bibfnamefont {T.}~\bibnamefont
  {Westerkamp}}, \bibinfo {author} {\bibfnamefont {M.}~\bibnamefont {Brando}},
  \bibinfo {author} {\bibfnamefont {N.}~\bibnamefont {Oeschler}}, \bibinfo
  {author} {\bibfnamefont {S.}~\bibnamefont {Wirth}}, \bibinfo {author}
  {\bibfnamefont {C.}~\bibnamefont {Krellner}}, \bibinfo {author}
  {\bibfnamefont {C.}~\bibnamefont {Geibel}},\ and\ \bibinfo {author}
  {\bibfnamefont {F.}~\bibnamefont {Steglich}},\ }\bibfield  {title} {\bibinfo
  {title} {Detaching the antiferromagnetic quantum critical point from the
  {{Fermi}}-surface reconstruction in {{YbRh$_2$Si$_2$}}},\ }\href
  {https://doi.org/10.1038/nphys1299} {\bibfield  {journal} {\bibinfo
  {journal} {Nat. Phys.}\ }\textbf {\bibinfo {volume} {5}},\ \bibinfo {pages}
  {465} (\bibinfo {year} {2009})}\BibitemShut {NoStop}%
\end{thebibliography}
%

\end{document}